\documentclass[11pt,a4paper]{paper}
\usepackage[utf8]{inputenc}
\usepackage[english]{babel}
\usepackage{amsmath}
\usepackage{amsfonts}
\usepackage{amssymb}
\usepackage{graphicx}
\usepackage[left=2cm,right=2cm,top=2cm,bottom=2cm]{geometry}
\usepackage{xcolor}
\usepackage[bookmarksopen=true]{hyperref}
\usepackage{bookmark}
\usepackage{bm}
\usepackage{comment}
\usepackage{multicol}
\usepackage{float}
\usepackage{color}
\usepackage{ulem}

\newlength{\bibitemsep}
\newlength{\bibparskip}
\let\oldthebibliography\thebibliography
\renewcommand\thebibliography[1]{%
  \oldthebibliography{#1}%
  \setlength{\parskip}{\bibitemsep}%
  \setlength{\itemsep}{\bibparskip}%
}

\begin{document}
\begin{center}
\large
\textbf{Effect of black hole--plasma system on light beams}\\

\hspace*{2cm}

\normalsize Matej S\'aren\'y{\footnote{matej.sareny@gmail.com}} \
and Vladim\'ir Balek{\footnote{balek@fmph.uniba.sk}}\\
\textit{Department of Theoretical Physics, Comenius University,
Bratislava, Slovakia}\\
\end{center}

\hspace*{1cm}

\section*{Abstract}
In the paper we discuss propagation of light around Kerr black
hole surrounded by non-magnetized cold plasma with infinite
conductivity. For that purpose, we use equations for propagation
of light rays obtained within Synge's approach in the
approximation of geometrical optics. We derive equation of
deviation of a ray propagating close to the reference ray, which
is a generalization of the well-known Jacobi equation, and use it
to calculate the modification of angular distribution of stars
observed close to a black hole surrounded by plasma, compared to
the uniform star distribution that would be seen without black
hole or plasma. We place the observer in the equatorial plane of
the Kerr black hole and try various choices of plasma
distributions described by mathematically simple formulae. Key
features of star distribution on a local sky near the black hole
are identified and the influence of plasma on them is discussed.

\noindent\textbf{Keywords}: relativistic geometric optics in
plasma, ray equation, ray deviation, star distribution function

\section{Introduction} The focus of our paper is the
behavior of light rays in the vicinity of a rotating uncharged
black hole (BH) in the presence of plasma. Since the discovery of
Kerr metric \cite{Kerr} in 1963, the behavior of light rays around
such BH has been intensively studied by many authors. Geodesics
around the Kerr BH were investigated for example in
\cite{Bardeen,Paganini} and detailed discussion of spherical
orbits can be found in \cite{Teo}. For an observer at arbitrary
position, some of the light rays are arriving ``from the BH'' (i.e.
no rays from distant stars can arrive from that direction), forming
the apparent ``shadow'' of the BH. The shape of the shadow was
investigated in \cite{Cunha2016,Cunha_review}. For an observer
close to the BH, the look of the night sky is distorted due to the
bending of the light rays. Such photographs were simulated in
\cite{Thorne,Kuchelmeister} using backward ray-tracing. Also, the
appearance of the accretion disc around the BH, as seen by the
observer outside the disc, was studied numerically
\cite{Thorne,Fukue,Luminet,Viergutz}.

The inclusion of optical effects of the plasma around the BH,
potentially present in the form of accretion disc, or a spherical
cloud of dilute material gathered from cosmic environment, can be
made by utilizing Synge's geometrical optics for an isotropic
medium with non-unit index of refraction \cite{Synge}. Lately,
this approach is being intensively applied in many papers. Effects
of plasma around a non-rotating BH were investigated in
\cite{Bisnovatyi,BKT2013,BKT2015,Bisnovatyi_3}, including changes
in deflection angle, shadow of the BH, magnification factor in
gravitational lensing etc., and additional effects arising in Kerr
metric were studied in \cite{Cunha_review, BKT2017,Kimpson,
Perlick}. In particular, the influence of plasma on the apparent
shadow of the BH was explored in \cite{Cunha_review, Perlick,
Huang}. As it turns out, among the properties of plasma only one
function, the density of electrons, has an effect on light rays,
while the 4-velocity of plasma drops out of the equations
entirely. Besides, the trajectory of photon depends on its
frequency due to the dispersive properties of plasma. The most
commonly used ``toy models'' for plasma distribution are
homogeneous plasma \cite{BKT2013,BKT2017, Perlick,Bisnovatyi_2,
Chakrabarty}, singular and nonsingular isothermal sphere
\cite{Bisnovatyi, Benavides-Gallego} and a sphere with power law
distribution \cite{Bisnovatyi,Bisnovatyi_3,BKT2017, Perlick,
Huang, Bisnovatyi_2, Chakrabarty, Benavides-Gallego, Rogers2015}.
Also, it would be interesting to consider disc-like distributions
where the disc's middle plane coincides with equatorial plane of
the Kerr BH. Realistic models of accretion discs are not commonly
used because of their intricate nature, involving discontinuities
of various types and uncertainty in the vertical profile. The
models also raise physical questions about the validity of
geometrical optics in a disc due to its optical thickness in most
cases, as well as its own radiation outshining any light rays from
distant stars that would propagate through it. The theoretical
work on accretion discs is reviewed in \cite{Abramowicz}; the most
popular models include thin discs
\cite{Shakura_Sunyaev,Novikov_Thorne, Page_Thorne, Penna},
advection dominated discs \cite{ADAF1,ADAF2} and Polish doughnuts
\cite{PolishDoughnut}.

The body of the paper is divided into four sections. In the second
section we rewrite Synge's equations into Lagrangian form,
obtaining equations of motion (EoMs) for the light ray in plasma
with such parametrization that ${\dot{x}}^\mu$ coincides with the
wave 4-vector. We build on this to derive ray deviation equation
(RDE), which is a generalization of equation of geodesic deviation
known from textbooks on general relativity. We also discuss the
conservation laws for the case of Kerr BH surrounded by stationary
axially-symmetrically distributed non-gravitating plasma. In the
third section we introduce the star distribution function (SDF),
which measures the angular dependence of star density as viewed by
an observer influenced by light bending in constrast to an
observer who sees homogeneous sky. In the fourth section we
develop tools for calculating the SDF numerically, following
techniques from \cite{Thorne,Pineault_Roeder_1,
Pineault_Roeder_2}. Furthermore, we discuss technical details of
our calculations, including observer specification, initial
conditions, plasma distribution ``toy models'' and some specifics
of our code. In the fifth section we discuss various features of
SDF as obtained in our calculations.

Throughout the paper we use the signature $(-+++)$ with Greek
spacetime indices and Latin spatial indices. Our other notations
follow the textbook \cite{Fecko}: we use the symbols $d$ for
exterior derivative, $\mathcal{L}$ for Lie derivative and $\nabla$
for the metric-compatible, torsion-free covariant derivative, and
adopt ``music notation'' $\flat$ and $\sharp$ for abstract
lowering and rising of indices. In index notation, these operators
are $(df)_\mu = f_{,\mu}$, $({\cal L}_{\bm U} \bm V)^\mu =
{V^\mu}_{,\nu} U^\nu - {U^\mu}_{,\nu} V^\nu$, $(\nabla_{\bm U} \bm
T)^{\rho \ldots \sigma}_{\mu \ldots \nu}= T^{\rho \ldots
\sigma}_{\mu \ldots \nu; \lambda} U^\lambda \equiv (T^{\rho \ldots
\sigma}_{\mu \ldots \nu, \lambda} + \Gamma^\rho_{\tau \lambda}
T^{\tau \ldots \sigma}_{\mu \ldots \nu} + \ldots -
\Gamma^\tau_{\mu \lambda} T^{\rho \ldots \sigma}_{\tau \ldots \nu}
- \ldots) U^\lambda$, $(\flat \bm V)_\mu = g_{\mu \nu} V^\nu$ and
$(\sharp \alpha)^\mu = g^{\mu \nu} \alpha_\nu$, where $f$ is
scalar, $\bm U$, $\bm V$ are vectors, $\alpha$ is covector, $\bm
T$ is tensor of an arbitrary rank and $\Gamma^\mu_{\nu \kappa}$
are coefficients of affine connection. (We restrict ourselves to
the expressions used in the article, therefore we do not write how
$d$ and $\cal L$ act on general tensors, just define their action
on scalars and vectors respectively.) We also utilize the
geometrized unit convention $G = c = 1$, and when considering Kerr
geometry we set $M$ (the mass of Kerr black hole) $=1$.

\section{Equations of motion for light rays in the presence of plasma}
\subsection{Equation of motion for a single ray}
Geometrical optics for isotropic media is developed in the book by
Synge \cite{Synge} in Hamilton -- Jacobi formalism. The basic
object is the eikonal $\sigma$, which satisfies the equation
\begin{eqnarray}\label{Eikonal_eqn}
\frac{1}{2}[g^{\mu \nu}+(1-n^2)V_m^{\mu}
V_m^{\nu}]\sigma_{,\mu}\sigma_{,\nu}=0
\end{eqnarray}
where $n$ is refraction index of the medium and $\bm V_m$ is its
4-velocity. In terms of the eikonal, the wave 4-vector at any
given point is defined as ${\bm k} = \sharp d\sigma$. For an
observer carrying a clock, a ruler and a protractor described by
an orthonormal tetrad $\bm e_{\hat {a}}$, the components
$k^{\hat{0}}=-\bm e_{\hat {0}}\cdot\bm k$ and $k^{\hat{i}}=\bm
e_{\hat {i}}\cdot\bm k$ give us the frequency and the wave vector
respectively, hence $\nabla_{{\bm V}_m} \sigma$ is minus the
frequency as observed in the medium rest frame.

The characteristics of the eikonal equation are called light rays
and are given by the Hamilton equations following from the
Hamiltonian
\begin{eqnarray}
H=\frac{1}{2}[\bm k^2+ (1-n^2)(\bm k\cdot \bm V_m)^2]
\end{eqnarray}
Since the Hamiltonian is independent on the parameter of the ray
$\lambda$, it must be constant, and from its definition it follows
that its only physically acceptable value is $H=0$.

The formula for the index of refraction of cold non-magnetized
plasma with infinite conductivity is (see for example
\cite{Feynman})
\begin{eqnarray}
n^2=1-\frac{\omega_{pl}^2}{\omega^2} \nonumber
\end{eqnarray}
The plasma frequency $\omega_{pl}$ can be expressed in terms of
the number density of free electrons measured in the plasma rest
frame $N(x)$ as $\omega^2_{pl}=e^2N(x)/(\varepsilon_0m_e)$, and as
seen from the formula for $n$, it places a lower limit on the
frequency $\omega$ of the wave measured in the plasma rest frame.
The resulting Hamiltonian for light rays in plasma is
\begin{eqnarray}\label{Hamiltonian_plasma}
H=\frac{1}{2}[\bm k^2+ \omega_{pl}^2(x)]
\end{eqnarray}
Thus, the light ray propagating in plasma behaves in the
approximation of geometrical optics as a particle with non-zero
rest mass that varies from point to point.

The form of the plasma Hamiltonian is familiar from analytical
mechanics and could have been obtained from the Lagrangian
$L=(1/2)[\bm k^2-\omega_{pl}^2(x)]$, hence the correct action for
a light ray is
\begin{eqnarray}\label{Ray_action_1}
S[x]=\frac{1}{2}\int_{\lambda_1}^{\lambda_2}[\bm k^2-\omega_{pl}^2(x)]d\lambda
\end{eqnarray}
where we have chosen the parameter $\lambda$ in such a way that
$\bm k= d/d\lambda$ (the parameter is not the proper time, the
parametrization ensures that $\bm k$ is wave 4-vector). Equation
of motion for a light ray is obtained from the condition $\delta
S=0$ for an arbitrary variation such that the boundary points
remain stationary. This yields
\begin{eqnarray}\label{Plasma_ray_eqn}
D\bm k=-\frac 12 \bm{\mathcal{A}}
\end{eqnarray}
where $\bm{\mathcal{A}}$ is the gradient of $\omega_{pl}^2$,
$\bm{\mathcal{A}}=\sharp d\omega_{pl}^2$, and $D=\nabla_{\bm k}$.
In index notation, the equation reads $Dk^\mu/d\lambda \equiv
k^\nu {k^\mu}_{;\nu} =-(1/2){\omega^2_{pl}}{}^{,\mu}$. As we can
see, the ray deviates from the geodesic in the direction of
maximal decrease of plasma density. The conservation law $H=const$
following from the absence of $\lambda$ in the Lagrangian allows
only for $H=0$ due to the eikonal equation. The law describes the
normalization of the wave 4-vector,
\begin{eqnarray}\label{4_vel_norm_1}
\bm k^2=-\omega_{pl}^2
\end{eqnarray}
In the special case with $\omega_{pl}=const$ we get geodesic
motion of a massive particle with constant rest mass. Cyclic
coordinates give rise to additional conservation laws.

The analogy between a massive particle and a photon propagating
through homogeneous plasma was noted in \cite{Bisnovatyi,Kulsrud}.
In the latter paper, the authors also show that in cold plasma the
transversal waves with dispersion relation (\ref{4_vel_norm_1})
are accompanied by a longitudinal wave whose frequency with
respect to the plasma rest frame is $\omega_{pl}$ irrespective of
the wave number.

For an arbitrary index of refraction, the ray equations can be
easily obtained by varying the Lagrangian $L = \bm k \cdot \bm U -
H$, $\bm U = d/d\lambda$, with respect to $\bm k$ and $x$. For the
Hamiltonian $H = (1/2) [\bm k^2 + (1 - n^2) \omega^2]$, $\omega =
-\bm k \cdot \bm V_m$, we find
\begin{eqnarray}
\begin{array}{ll}
\bm U &= \bm k-Q {\bm V}_m\\
\nabla_{\bm U} \bm k &= Q\bm W(\bm k, .) + \bm q,\\
\end{array} \nonumber
\end{eqnarray}
where $Q = (1 - n^2)\omega - n(\partial_\omega n)_x\omega^2$, $\bm
q = n(\sharp dn)_\omega \omega^2$ and $\bm W = \sharp \nabla \flat
{\bm V}_m$ (in index notation, ${W_\mu}^\nu =
{(V_m)_\mu}^{;\nu}$). For cold plasma $Q = 0$ and $\bm q= -
\bm{\mathcal A}/2$, hence we arrive once again at equation
(\ref{Plasma_ray_eqn}).

\subsection{Equation of motion for infinitesimally close rays}
\subsubsection*{Geodesic deviation}
A powerful tool to describe a light beam (a bundle of nearby light
rays) in vacuum is the Jacobi equation a.k.a. equation of geodesic
deviation
\begin{eqnarray}
D^2\bm\xi=R(\bm k, \bm\xi)\bm k\nonumber
\end{eqnarray}
The equation describes evolution of the connecting vector $\bm\xi$
pointing from the reference geodesic to a geodesic infinitesimally
close to it. The vector $\bm k$ is the wave 4-vector, however, not
timelike as before but null, $\bm k^2=0$, and $R$ is the curvature
operator, $R(\bm A,\bm B)=\nabla_{\bm A}\nabla_{\bm B}-\nabla_{\bm
B}\nabla_{\bm A}-\nabla_{[\bm A, \bm B]}$. The derivation of this
equation can be found for example in \cite{MTW} or \cite{Fecko}.
The connecting vector is supposed to be Lie-transported along the
reference geodesic, $\mathcal{L}_{\bm k}\bm\xi\equiv[\bm k, \bm
\xi]=0$. The advantage of the Jacobi equation is that it needs to
be integrated only along the reference geodesic, but we gain also
information about the neighboring geodesics from it.

\subsubsection*{Ray deviation}
Let us consider the action for rays in plasma
(\ref{Ray_action_1}), but for a nearby ray
$y^\mu=x^\mu+\varepsilon\xi^\mu$. The generalization of the Jacobi
equation in the presence of plasma reads
\begin{eqnarray}\label{Ray_deviation_equation}
D^2\bm\xi=R(\bm k,\bm\xi)\bm k-\frac{1}{2}\sharp\bm{\mathcal{H}}(\bm\xi,.)
\end{eqnarray}
where $\bm{\mathcal{H}}$ is the covariant Hessian matrix of
$\omega_{pl}^2$, $\bm{\mathcal{H}}=\nabla d\omega_{pl}^2$; or in
components, ${\mathcal{H}}_{\mu\nu}=(\omega^2_{pl})_{;\mu\nu}$.
The derivation of this equation is discussed in the appendix
\ref{app:RDE}. The conservation law due to the absence of
$\lambda$ in the Lagrangian governing the evolution of $\xi$
yields
\begin{eqnarray}\label{4_vel_norm_2}
\bm k\cdot D\bm\xi+\frac{1}{2}\bm\xi\cdot\bm{\mathcal{A}} = 0
\end{eqnarray}
Actually, the conservation law implies just that the expression on
the left hand side is constant, but the differentiation of the
constraint (\ref{4_vel_norm_1}) restricts the constant to be zero.
If we choose an arbitrary observer with 4-velocity $\bm V$, we can
decompose Jacobi vector into its spatial and time component with
respect to that observer, and assuming $\bm V\neq C\bm k$, the
spatial part can further be decomposed into longitudinal and
transverse part w.r.t. the directional vector of the reference ray
$\bm k$ (or alternatively, the spatial part of this vector w.r.t.
$\bm V$). The transverse part can be, under certain conditions,
used to measure the thickness of a narrow light beam.

Light bundles in cold plasma were investigated also in [39]. In
the approach developed there, light rays are mapped onto timelike
geodesics in a conformally rescaled metric $\tilde {\bm g} =
\omega_{pl}^2 \bm g$, in which the connecting vector
$\bm{\tilde{\xi}}$ obeys standard Jacobi equation (with the
4-velocity normalized to unity in the rescaled metric and the
connecting vector Lie-transported along it, i.e. satisfying
$[\bm{\tilde{k}}, \bm{\tilde{\xi}}]=0$). The authors derive from
the equation for $\bm{\tilde{\xi}}$ equations for optical
parameters, describing the beam in a basis parallel transported in
the conformal metric (Sachs basis). We, on the other hand, use for
the description of the beam the connecting vector $\bm \xi$ which
we determine directly from equation
(\ref{Ray_deviation_equation}). The two vectors differ, because
the parameters in the two theories, with the metrics $\bm g$ and
$\tilde {\bm g}$, differ: parameter $\lambda$ in the former theory
is related to parameter $\tilde \lambda$ in the latter theory by
the formula $d\tilde \lambda = \omega_{pl}^2 d\lambda$. As a
result, vector $\tilde {\bm \xi}$ can be expressed in terms of
vector $\bm \xi$ as
\begin{eqnarray}
\tilde {\bm \xi} = \bm \xi + \Phi \bm k \qquad\qquad\qquad \Phi =
-\omega_{pl}^{-2} \int \bm{\mathcal{A}} \cdot \bm \xi d\lambda
\nonumber
\end{eqnarray}

\subsubsection*{Evolution of a narrow elliptical light beam}
Consider a light beam with elliptical cross-section. EoMs for a
reference ray and an adjacent ray comprise a system of 8 second
order ordinary differential equations (ODE), or equivalently 16 first
order ODE
\begin{eqnarray}\label{EoM_abstract}
\begin{array}{ll}
Dx & =\bm k\\
D\bm k & =-(1/2)\bm{\mathcal{A}}\\
\end{array}
\qquad\qquad\qquad
\begin{array}{ll}
D\bm\xi & =\bm{\mathcal{V}}\\
D\bm{\mathcal{V}} & =R(\bm k,\bm\xi)\bm k-(1/2)\sharp\bm{\mathcal{H}}(\bm\xi,.)
\end{array}
\end{eqnarray}
Equation of ellipse can be written as
$\vec{r}=\vec{a}\cos\sigma+\vec{b}\sin\sigma$ with general vectors
$\vec{a}$ and $\vec{b}$ (they can be made orthogonal by shifting
the parameter $\sigma$ by the constant $\rho$ such that
$\tan(2\rho)=2\vec{a}\cdot\vec{b}/(a^2-b^2)$). To evolve the whole
beam along the reference ray we plug this Ansatz into the
equations for ray deviation, i. e. for an arbitrary point at the
boundary of the beam we write $\bm\xi=\bm \xi_{(1)}\cos\sigma+\bm
\xi_{(2)}\sin\sigma$, $\bm{\mathcal{V}} = \bm{\mathcal{V}}_{(1)}
\cos\sigma + \bm{\mathcal{V}}_{(2)}\sin\sigma$ and demand the EoMs
to be valid for every $\sigma$. As a result we obtain the
equations for two ellipse-generating adjacent rays ($k=1,2$):
\begin{eqnarray}
\begin{array}{ll}
\qquad D\bm
\xi_{(k)}=\bm{\mathcal{V}}_{(k)}\qquad\qquad\qquad\qquad
D\bm{\mathcal{V}}_{(k)}=R(\bm k,\bm \xi_{(k)})\bm
k-(1/2)\sharp\bm{\mathcal{H}}(\bm \xi_{(k)},.)
\end{array}\nonumber
\end{eqnarray}
Thus, determining the form of a beam requires integration of 24
equations (if we had at the beginning a beam with elliptical cross
section).

\subsection{Conserved quantities}
Kerr geometry has two cyclic coordinates $t$, $\varphi$,
associated with Killing vectors $\partial_t$, $\partial_\varphi$.
As a result, in a stationary, axially symmetric plasma
distribution where $N = N(r,\vartheta)$ we obtain two conserved
quantities for a light ray described by the action
(\ref{Ray_action_1}),
\begin{eqnarray}\label{Ray_cons_laws_1}
\begin{array}{ll}
k_t&=\partial L/\partial\dot{t}=-E\\
k_\varphi&=\partial L/\partial\dot{\varphi}=\mathcal{L}
\end{array}
\end{eqnarray}
These suggestively denoted constants represent (up to a
multiplicative constant with the dimension of $\hbar$) energy and
angular momentum w.r.t. the axis of symmetry, as measured by the
observer at rest at infinity. Additional conservation laws can be
derived from the two-ray action (\ref{Ray_action_2}), assuming
once again stationary axially symmetric plasma distribution. The
cyclic coordinates are explicitly visible when the Lagrangian
appearing in the action is written in the coordinate frame:
\begin{eqnarray}
\mathcal{L}=g_{\mu\nu}\dot
x^{\mu}\dot\xi^{\nu}+\Gamma_{\mu\nu\rho}\dot x^{\mu}\dot
x^{\nu}\xi^\rho-\frac{1}{2}\xi^\mu \mathcal{A}_\mu\nonumber
\end{eqnarray}
The four coordinates that are absent here, hence are cyclic, are
$t,\varphi,\xi^t,\xi^\varphi$. The second pair yields the
conservation of $k_t, k_\varphi$ again, while the first pair
results in additional conserved quantities
\begin{eqnarray}\label{Ray_cons_laws_2}
\begin{array}{llll}
P_t&=\partial \mathcal{L}/\partial \dot t&=(D\xi)_t+ \Gamma_{\mu t
\nu}k^\mu\xi^\nu &=\bm{\mathcal{V}}\cdot\partial_t+\bm
k\cdot\nabla_{\bm\xi}
\partial_t\\
P_\varphi &=\partial \mathcal{L}/\partial \dot
\varphi&=(D\xi)_\varphi+ \Gamma_{\mu \varphi \nu}k^\mu\xi^\nu
&=\bm{\mathcal{V}}\cdot\partial_\varphi+\bm k\cdot
\nabla_{\bm\xi}\partial_\varphi
\end{array}
\end{eqnarray}
The coordinate-free formula is convenient for recalculation of the
conserved quantities into an arbitrary basis. Note that the
existence of the second pair of conserved quantities comes from
the fact that we have two conserved quantities per ray.

\section{Jacobi map from the local sky to the celestial sphere}
As an example let us choose the Kerr BH surrounded by plasma with
negligible gravitational influence. Relevant formulae for Kerr
geometry are found in appendix \ref{app:Kerr}. Consider an
observer outside of the event horizon, characterized by their
position $P=(r_0, \vartheta_0,\varphi_0)$ and 4-velocity $\bm V$;
moreover, suppose they can observe the radiation only in a narrow
band of frequencies centered around $\omega_{obs}$. The set of all
directions in which the observer can look is a unit sphere $S^2$,
which may be called the ``local sky'' of the observer. The stars
located far away from both the observer and the black hole can be
thought of as living on a sphere with large radius which we will
call the ``celestial sphere''. Denote the coordinates on the local
sky $(\bar\vartheta, \bar\varphi)$ and the coordinates on the
celestial sphere $(\vartheta_\infty,\varphi_\infty)$ and define
Jacobi map $\mathcal{J}$ that assigns a point on the local sky to
a point on the celestial sphere by evolving a ray with initial
frequency $\omega_{obs}$, which arrives to the observer from the
direction given by the angular coordinates $(\bar\vartheta,
\bar\varphi)$, back in time according to the evolution equation
(\ref{Plasma_ray_eqn}):
\begin{eqnarray}
\mathcal{J}(\omega_{obs}, P, \bm{V}): (\bar{\vartheta},\bar\varphi)\mapsto (\vartheta(\lambda=-\infty),
\varphi(\lambda=-\infty))\nonumber
\end{eqnarray}
The angles $\vartheta,\varphi$ appearing here are Boyer-Lindquist
coordinates of the evolved ray. The definition only works if the
light ray actually reaches infinity; if it does not for some
reason, e.g. because it arrives ``from the BH'', the corresponding
point of Jacobi map remains undefined.

Suppose the stars are distributed uniformly on the celestial
sphere. Consequently, the number of stars seen in the solid angle
$d\Omega_\infty$ is $dN=(N_s/4\pi)d\Omega_\infty$, where $N_s$ is
the total number of stars on the night sky. On the local sky, the
distribution of stars will not be uniform due to the bending of
the light rays. Thus, it will have the form
\begin{eqnarray}
dN=\frac{N_s}{4\pi}n_s(\bar\vartheta, \bar\varphi)d\bar\Omega\nonumber
\end{eqnarray}
The formula defines the function $n_s$ -- the star distribution
function (SDF) representing relative increase in the angular
density of stars in comparison to the uniform case. Rewriting $dN$
in terms of $d\Omega_\infty$ we obtain
\begin{eqnarray}\label{SDF_def}
n_s=\frac{d\Omega_\infty}{d\bar\Omega}
\end{eqnarray}
In the empty space without plasma, the SDF is identically equal to
one. Using the Jacobi map we can relate the two solid angles to
each other,
\begin{eqnarray}
d\Omega_\infty = \sin\vartheta_\infty d\vartheta_\infty
d\varphi_\infty = \sin\vartheta_\infty \vert J\vert d\bar
\vartheta d\bar\varphi
=\frac{\sin\vartheta_\infty}{\sin\bar\vartheta}\vert J\vert
d\bar\Omega\nonumber
\end{eqnarray}
where $J$ is the Jacobian of the Jacobi map. Therefore, the SDF in
the presence of a black hole and plasma can be calculated as
\begin{eqnarray}\label{SDF_jacobian_formula}
n_s=\frac{\sin\vartheta_\infty}{\sin\bar\vartheta}\left \vert
\frac{\partial \vartheta_\infty}{\partial\bar\vartheta}
\frac{\partial \varphi_\infty}{\partial\bar\varphi}-\frac{\partial
\varphi_\infty}{\partial\bar\vartheta} \frac{\partial
\vartheta_\infty}{\partial\bar\varphi}\right\vert
\end{eqnarray}
However, this formula is not very suitable for numerical
calculations. First, it contains derivatives which are complicated
to calculate numerically, since that would involve many
integrations of the ray evolution equation. Second, it contains
the sine in the denominator which can be equal to zero. The RDE
offers also another way to calculate the SDF, as we will see
later.

\section{Numerical computation of SDF using LNRF}
The following recipe is a modification of the procedure used by
Thorne et al. \cite{Thorne}, when calculating shots for the
movie Interstellar. We have modified the procedure to allow for a
non-unit refraction index. Also, we chose to do the calculations
in the locally non-rotating reference frame (LNRF), rather than
adopting phasors (complex numbers) and adapted reference frame
used there. The goal is to find the SDF in Kerr geometry with
various plasma distributions. We have written a program
integrating previously described equations of ray deviation to
calculate the SDF. The program is written in Python 3 and uses the
integrator \textit{lsode} for ODE from
the package \textit{scipy.integrate.ode} \cite{ode}. Some SDFs
obtained by this method can be found in Appendix
\ref{app:results}.

\subsection{Specifics of the program}
We have numerically calculated the SDFs as seen by an observer
orbiting the Kerr black hole in the equatorial plane, possibly in
the presence of plasma. We have left the following parameters
free: the Kerr parameter $a$, the radial coordinate of the
observer $r_0$, the orbiting velocity of the observer as measured
in the LNRF $B$, the observational frequency as measured by the
observer $\omega_P$, and the plasma distribution
$\omega_{pl}^2(r,\vartheta)$ (we assumed it is smooth enough, so
that the light does not refract by a finite angle after passing an
infinitesimal distance along its path). Due to the mirror symmetry
w.r.t the equatorial plane, it was sufficient to calculate the
star distribution in the lower hemisphere w.r.t. the equatorial
plane (the surface $\bar{\varphi}\in [0,\pi]$, see figure
\ref{fig:obs_angles}).

At the observer's local sky, we have set up a grid of angles from
the interval $(\bar{\vartheta},\bar{\varphi})\in
[0,\pi]\times[0,\pi]$. Due to the method of calculation used, the
accuracy of the values obtained at individual elements of the grid
was independent from the size of the grid, which determined only
the resolution of the final figure. In our calculation we could
set the size of the grid independently in $\bar{\vartheta}$ and
$\bar{\varphi}$ directions. To obtain the SDF for each point of the grid, we have numerically
integrated the RDE (twice) and then computed
the value of the SDF using the formula (\ref{SDF_num_formula}).
The numerical integration was the most time-consuming process,
therefore the calculation time has increased roughly linearly with
the grid size. To decide when (if at all) the ray reached
infinity, we have defined the ``numerical infinity'' at
$r_{\infty}=10,000$. We have also set the maximum value of the
affine parameter allowed for a ray to $\lambda_{max} = 5\times 10^5
\omega_P^{-1}$. Rays that did not reach $r_\infty$ before the
parameter has assumed this value, or entered the event horizon in
the meantime, were assigned blue color in the final picture
(meaning undefined $n_s$). Note that in vacuum there also exists a
more sophisticated criterion determining when the rays escape to
infinity, which can be found e.g. in \cite{Thorne}.

We have displayed the SDFs in figures consisting of two circles,
which represent the ``inner'' and ``outer'' hemisphere oriented
towards and away from the black hole respectively. The values of
SDF were color-coded into a hybrid linear ($n_S\in [0,3]$) and segmented ($n_S>3$) colormap, since
SDF tends to diverge when approaching the BH's shadow and the
resulting large values in a small region around the shadow would
mess up the color-coding of the rest of the picture on a purely linear colormap.

\subsection{Initial conditions}
\subsubsection*{Initial conditions for the reference ray}
 Consider an  observer at an event $P$ orbiting in the
equatorial plane of the Kerr black hole with the velocity with
respect to the LNRF $-1<B<1$, and introduce two orthonormal bases
along the observer's worldline, the LNRF basis $(\bm e_{\hat{t}},
\bm e_{\hat{r}}, \bm e_{\hat{\vartheta}}, \bm e_{\hat{\varphi}})$
and the observer basis $(\bm E_t = \bm V,\bm E_r,\bm
E_\vartheta,\bm E_\varphi)$. The components of the 4-velocity of
the observer $\bm{V}$ as measured in the LNRF are given by the
special relativistic formula $V^{\hat{\mu}}=\gamma(B)(1,0,0,B)$,
where $\gamma(B)=(1-B^2)^{-1/2}$ is the Lorentz factor.
Correspondingly, $e_{\hat{t}}^a=\gamma(B)(1,0,0,-B)$, which
reflects the fact that the LNRF moves w.r.t. the observer $P$ with
the velocity $-B$. The complete set of transformation relations is
given by the boost formulae (perpendicular {vectors do not change
and the fourth vector is just a normalized combination of the
original vectors orthogonal to the first one):
\begin{eqnarray}\label{Obs_to_lnrf}
e_{\hat{t}}=\gamma(B)(E_t-BE_\varphi) \quad\quad e_{\hat{r}}=E_r
\quad\quad  e_{\hat{\vartheta}}=E_\vartheta \quad\quad
e_{\hat{\varphi}}=\gamma(B)(-BE_t+E_\varphi)
\end{eqnarray}
In the instantaneous physical 3-space of the observer $P$,
standard spherical coordinates $(\bar{\vartheta},\bar{\varphi})$
are introduced using the correspondence
$(E_\varphi,E_\vartheta,-E_r)\leftrightarrow(\epsilon_x,\epsilon_y,\epsilon_z)$
(the zenith angle $\bar{\vartheta}$ is measured from the radial direction
pointing to the black hole and the azimuthal angle $\bar{\varphi}$ is
measured from the vector $E_\varphi$ in the clockwise direction w.r.t. $E_r$, see figure \ref{fig:obs_angles}).

\begin{figure}[h]
\includegraphics[scale=1]{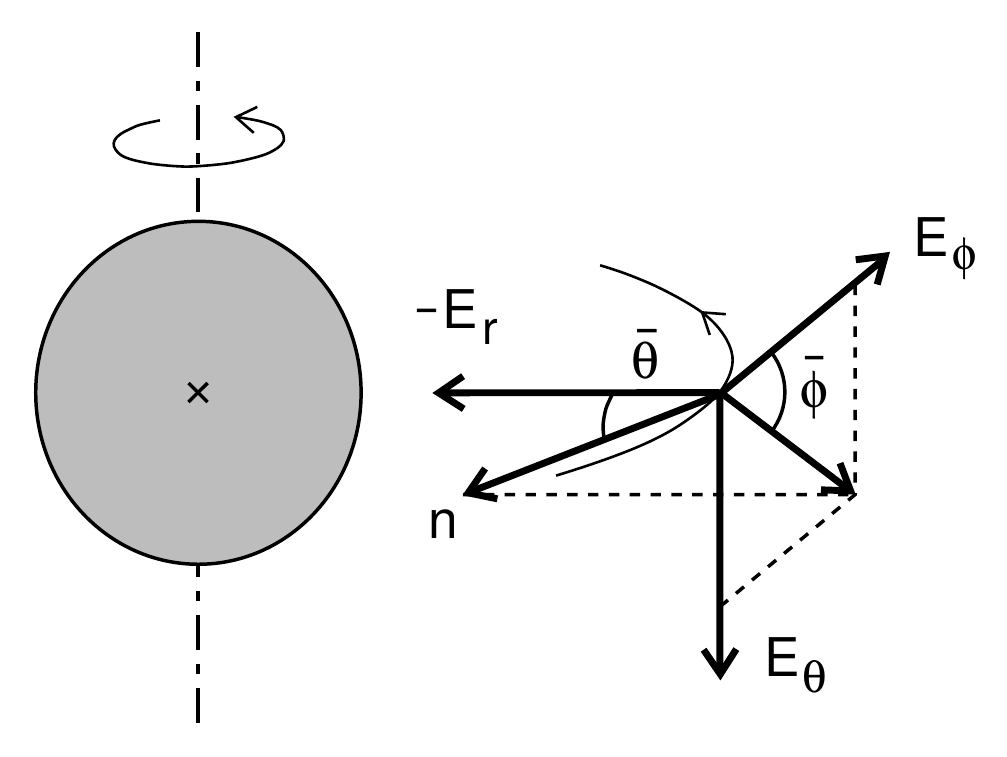}
\centering \caption{Definition of observer's angles
$\bar\vartheta,\bar\varphi$} \label{fig:obs_angles}
\end{figure}

In the observer's frame, the wave 4-vector of the reference ray is
initialized as $k^a=\omega_P(1,-n_P\vec{n})$, where
$\vec{n}=(-c_{\bar{\vartheta}},
s_{\bar{\vartheta}}s_{\bar{\varphi}},
s_{\bar{\vartheta}}c_{\bar{\varphi}})$ is the unit vector in the
direction in which the observer is looking (the photon is heading
in the opposite direction, hence the extra minus sign) and $n_P=\sqrt{1-\omega_{pl}^2/\omega_P^2}$ is the index of refraction
at the point where the observer is located. After initialization,
we have to integrate towards negative values of the affine
parameter (into the past).

\subsubsection*{Initial conditions for the beam}
All rays of the beam are focused at the point $P$ at $\lambda=0$
and spread into a circular beam, seen by the observer within the
solid angle $d\Omega_P$ at $\lambda=-d\lambda$. Focusing of the
beam implies initial conditions (ICs) $\bm \xi_{(k)}(0)=0$ and
spreading of the beam into the solid angle $d\Omega_P$ suggests
$\bm {\mathcal{V}}_{(k)}(0)=\bm{\dot{\xi}}_{(k)}(0)$ with $\bm
\xi(-d\lambda)=-(\bm {\mathcal{V}}_{(1)}(0)\cos\sigma+ \bm
{\mathcal{V}}_{(2)}(0)\sin\sigma)d\lambda$. The spatial part of
the 4-vectors $\bm {\mathcal{V}}_{(k)}$ is supposed to be two
independent vectors $\vec V_{(k)}$ perpendicular to $\vec{n}$ and
to each other (in the sense of the induced 3-metric w.r.t. $\bm
V_P$), which have both the magnitude $R$, therefore the radius of
the circle as measured from $P$ is\footnote{$\epsilon$ comes from
our convention that the connecting vector points to an
infinitesimally close point, $y^\mu = x^\mu + \epsilon \xi^\mu$}
$\epsilon Rd\lambda$. The radius of the circle enters ICs via $\pi
\epsilon^2R^2d\lambda^2=dr^2d\Omega_P=\omega_P^2n_P^2d\lambda^2d\Omega_P$,
therefore $R=\epsilon^{-1}\omega_Pn_P\sqrt{d\Omega_P/\pi}$. The time
component of the vector $\bm {\mathcal{V}}$ is given by the
condition $\bm k\cdot\bm{\mathcal{V}}=0$ (see
(\ref{4_vel_norm_2})), and it vanishes,
${\mathcal{V}}^t_{(k)}(0)=0$, for our choice of spatial ICs.

\subsubsection*{Initialization of differential equations}
At the point $P$, 10 dynamic variables entering the evolution
equations must be initialized together with 3 required conserved
quantities. Position of the central ray is initialized by the
choice of the point $P$ somewhere in the equatorial plane (and
above the event horizon), $(r_0$, $\vartheta_0)=(r_P,\pi/2)$.
Vector $\bm\xi$ is initially zero in the frame of the observer $P$
with our choice of ICs, therefore it is zero in any other frame
too. The wave 4-vector, for both the reference ray and the
adjacent ray, needs to be transformed from the observer's frame to
LNRF, which can be done by using inverse transformation to the
transformation (\ref{Obs_to_lnrf}). To determine the conserving
constants we must further pass from LNRF to the coordinate basis;
for that purpose, use (\ref{lnrf_to_coordinate}) and the fact that
covector components transform like basis vectors. Finally, notice
that $P_\varphi=\mathcal{V}_{\varphi 0}$ thanks to our choice of
ICs. Below we give the full list of ICs expressed in terms of the
parameters $\omega_P,\bar{\varphi}$, $\bar{\vartheta}$, $R=
\epsilon^{-1} \omega_Pn_P\sqrt{d\Omega_P/\pi}$ and the two
ellipse-generating vectors $\vec{b}_{(k)}$, which are unit
3-vectors perpendicular to
$\vec{n}(\bar{\vartheta},\bar{\varphi})$ (they can also be chosen
to be perpendicular to each other,
$\vec{b}_{(1)}\cdot\vec{b}_{(2)}=0$):
\begin{equation}\label{IC_lnrf}
\begin{array}{ll}
r_0&=r_P\\
\vartheta_0&=\pi/2\\
k^{\hat{r}}_0&=\omega_Pn_Pc_{\bar{\vartheta}}\\
k^{\hat{\vartheta}}_0&=-\omega_Pn_Ps_{\bar{\vartheta}}s_{\bar{\varphi}}\\
k^{\hat{t}}_0&=\gamma(B)\omega_P[1-Bn_Ps_{\bar{\vartheta}}c_{\bar{\varphi}}]\\
k^{\hat{\varphi}}_0&=\gamma(B)\omega_P[B-n_Ps_{\bar{\vartheta}}c_{\bar{\varphi}}]
\end{array}
\qquad\qquad\qquad
\begin{array}{ll}
\xi^{\hat{t}}_0&=\xi^{\hat{r}}_0=\xi^{\hat{\vartheta}}_0=\xi^{\hat{\varphi}}_0=0\\
\mathcal{V}^{\hat{r}}_0&=Rb^{\tilde{r}}\\
\mathcal{V}^{\hat{\vartheta}}_0&=Rb^{\tilde{\vartheta}}\\
P_\varphi&=\bar{\omega}\gamma(B)Rb^{\tilde{\varphi}}\\
\varepsilon&=\alpha k^{\hat{t}}_0+\omega\bar{\omega}k^{\hat{\varphi}}_0\\
\mathcal{L}&=\bar{\omega}k^{\hat{\varphi}}_0\\
\end{array}
\end{equation}
where $\alpha$, $\omega$ and $\bar{\omega}$ are defined in the
appendix \ref{app:Kerr}. Note that the EoMs (\ref{EoM_lnrf}) are
invariant w.r.t. the rescaling $x\mapsto x$, $\bm k\mapsto K\bm
k$, $\omega_{pl}\mapsto K\omega_{pl}$, $\lambda\mapsto
K^{-1}\lambda$, $\bm\xi\mapsto Q\bm\xi$, $\bm{\mathcal{V}}\mapsto
KQ\bm{\mathcal{V}}$. This means that we can freely choose the
``typical scale'' for $\bm k$ and $\bm\xi$ without the necessity
to change the equations in any way. A good choice is $K=\omega_P$
and $Q=R/\omega_P$, since it removes the superfluous factors from
the ICs.

\subsection{Evaluation of SDF}
Provided the plasma density gradient falls to zero sufficiently
fast when $r\rightarrow\infty$, the reference ray 4-velocity at
$\lambda=-\infty$ equals the wave 4-vector of radially incoming
ray, $k^{\hat{\mu}}=\omega_\infty(1,-n_\infty,0,0)$. The
cross-sectional area of the beam is therefore measured in the
$Span(\bm e_{\hat{\vartheta}},\bm e_{\hat{\varphi}})$ plane. Using
the Thorne's approach \cite{Thorne} we write the major and minor
angular semiaxis of the ellipse as $\delta_\pm=-dY_\pm/dr$, where
$Y_\pm$ is the major and minor linear semiaxis of the ellipse.
Since $\bm \xi_{(1)}$ is in general not perpendicular to $\bm
\xi_{(2)}$, it is more convenient (following \cite{Thorne}) to use the description
in which the ellipse is encoded in the complex variable
$\zeta=\xi^{\hat{\vartheta}}+i\xi^{\hat{\varphi}}$. It holds
\begin{eqnarray}
\zeta=\frac{1}{2}\left\lbrace \left[(\xi_{(2)}^{\hat{\varphi}}+
\xi_{(1)}^{\hat{\vartheta}})+i(\xi_{(1)}^{\hat{\varphi}}-
\xi_{(2)}^{\hat{\vartheta}})\right]e^{i\sigma}+
\left[(\xi_{(1)}^{\hat{\vartheta}}-\xi_{(2)}^{\hat{\varphi}}
)+i(\xi_{(1)}^{\hat{\varphi}}+\xi_{(2)}^{\hat{\vartheta}})\right]
e^{-i\sigma}\right\rbrace\equiv\frac{1}{2}( \kappa e^{i\sigma}+
\mu e^{-i\sigma})\nonumber
\end{eqnarray}
and, consequently, $Y_\pm=\frac{1}{2}
\vert\sqrt{\kappa\kappa^*}\pm\sqrt{\mu\mu^*}\vert$ (the major and
minor semiaxis arises as the constructive and destructive sum of
the phasors respectively). Also, if we notice that at infinity
$\xi_{(k)}^{\hat{\mu}}\approx {\mathcal{V}}_{(k)}^{\hat{\mu}}
\lambda$ and $dr\approx -\omega_\infty n_\infty d\lambda$, we find
\begin{eqnarray}
\delta_\pm=\frac{1}{2}\frac{1}{n_\infty\omega_\infty}\vert
\sqrt{\kappa_\mathcal{V}\kappa_\mathcal{V}^*}\pm\sqrt{\mu_\mathcal{V}
\mu_\mathcal{V}^*}\vert\nonumber
\end{eqnarray}
where the subscript $\mathcal{V}$ means that we have to replace
all $\bm\xi$'s with $\bm{\mathcal{V}}$'s in the definitions of the
corresponding variables. The last step is evaluating
$d\Omega_\infty=\pi\delta_+\delta_-$. From the expression for
$\delta_\pm$ it follows
\begin{eqnarray}
d\Omega_\infty=\frac{\pi}{n_\infty^2\omega_\infty^2}
\vert{\mathcal{V}}_{(1)}^{\hat{\vartheta}}{\mathcal{V}}_{(2)}^{\hat{\varphi}}
-{\mathcal{V}}_{(2)}^{\hat{\vartheta}}{\mathcal{V}}_{(1)}^{\hat{\varphi}}
\vert\nonumber
\end{eqnarray}
and if we insert $d\bar\Omega=d\Omega_P$ into the definition of
SDF and denote $\tilde{\mathcal{V}}=\mathcal{V}/R$, we obtain
\begin{eqnarray}\label{SDF_num_formula}
n_s=\frac{n_P^2\omega_P^2}{n_\infty^2\omega_\infty^2}
\vert{\tilde{\mathcal{V}}}_{(1)}^{\hat{\vartheta}}
{\tilde{\mathcal{V}}}_{(2)}^{\hat{\varphi}}-
{\tilde{\mathcal{V}}}_{(2)}^{\hat{\vartheta}}
{\tilde{\mathcal{V}}}_{(1)}^{\hat{\varphi}}\vert
\end{eqnarray}

\subsection{Plasma distribution}
In our calculations we have used various ``toy models'' of the
plasma distribution, all of which can be found in the literature
except for the last one which we have proposed ourselves. The
advantage of these models is their smoothness, as well as the fact
that the derivatives of the plasma density can be expressed
analytically. The most trivial case is vacuum with $\rho=0$.
Another simple case is the distribution $\rho=const$, when the ray
equation coincides with the equation for a massive particle in
vacuum. A model often used in the literature \cite{Bisnovatyi,
Benavides-Gallego} is the non-singular isothermal sphere (NSIS),
defined as
\begin{eqnarray}\label{Non-sing_is_sph}
\rho=\frac{K}{r^2+r_c^2}
\end{eqnarray}
where $K$ and $r_c$ are positive constants. The same model with
$r_c=0$ is called the singular isothermal sphere \cite{Bisnovatyi,
Benavides-Gallego}. In Newtonian gravity, such sphere can be
obtained as a solution for a static non-rotating self-gravitating
gas cloud with the same temperature of the gas everywhere,
supported against the collapse solely by the pressure gradient.
Finally, as a disc-like distribution we have used a non-singular
isothermal sphere modified by a function that flattens the
distribution towards the equatorial plane,
\begin{eqnarray}\label{Disclike_NSIS}
\rho=\frac{K}{r^2+r_c^2}\exp
\left[-\frac{(\vartheta-\pi/2)^2}{s^2}\right]
\end{eqnarray}
where the parameter $s$ determines the magnitude of flattening.

\section{General features of SDFs}
We have investigated four plasma distributions, namely no plasma,
homogeneous plasma, NSIS (distribution (\ref{Non-sing_is_sph}))
and flattened NSIS (distribution (\ref{Disclike_NSIS})). The
4-velocity of the plasma was not important since it vanishes from
the equations. We have chosen the locally non-rotating family of
observers, since they filter out the special relativistic effects
as much as possible and exist also near the event horizon
(contrary to the static observers who become tachyonic inside
ergosphere). In addition to that, we have limited ourselves to the
observers located in the equatorial plane. The main properties of
the SDFs obtained in this way are:

a) no plasma: it is well known that purely gravitational bending
of light is achromatic (frequency independent), hence the SDF is
independent of $\omega_{obs}$ in the case with no plasma. Far from
the black hole, the SDF is close to one in almost all directions,
it changes rapidly only in the vicinity of the black hole shadow,
decreasing to zero at Einstein's ring and diverging at
the edge of the shadow. The steep rise of the SDF near the shadow
is caused by the multiple image generation by the BH. In fact, the SDF oscillates between zero and large values infinitely many times when approaching the shadow, but in the calculations presented here we do not see this behaviour  due to low resolution. As we approach the BH, the
behavior of the SDF stays qualitatively the same, but its features
become more pronounced. Non-zero spin of the BH introduces
asymmetry in SDF w.r.t. the azimuthal direction, with slightly
greater values of SDF in the prograde direction. In particular,
near the BH the outward-facing maximum of SDF is visibly shifted
from $\bar\vartheta=\pi$ in the prograde direction if the BH is
rotating (see fig. \ref{fig:vac}). As the observer approaches the event horizon, the BH
shadow becomes ever more pronounced, eventually encompassing the
whole sky.

b) homogeneous plasma: in this case, the light rays behave as
massive test particles in Kerr geometry. Recalling that
$k^{\hat{\mu}}=\omega_{obs}(1,-n_{obs}\vec{n})$, we see
that $n_{obs}$ plays the role of launching velocity and
$\omega_{obs}$ is, modulo factor $1/\omega_{pl}$, the launching
Lorentz factor. If at least some rays are to reach infinity, the
launching velocity must exceed the escape velocity, $\omega_{obs}>
\omega_{min}> \omega_{pl}(r_{obs})$. Thus, the SDF acquires a new
property: the night sky is not visible below $\omega_{min}$. In
case of rotating BH or a boosted frame, the escape energy exhibits
non-trivial dependence on the launching direction. Finally, in the
limit of large launching velocity, $\omega_{obs} \gg
\omega_{pl}(r_{obs})$, we regain the vacuum picture.

c) non-singular isothermal sphere: the $r$-dependence of NSIS
leads to radial acceleration, pushing the light rays away from the
BH (see fig. \ref{fig:vertical_raytrace} c)). This fights the
gravitational pull of the black hole and may result in erasing the
black hole shadow from the night sky entirely.

d) flattened NSIS: the disc-like shape of this distribution causes
an additional deviation of the acceleration of rays from radial
direction: the rays are pushed away from the equatorial plane (see
fig. \ref{fig:vertical_raytrace} d)). As a consequence, the number
of rays the observer receives when looking in equatorial plane
directions is hugely enhanced and SDF assumes large values in
these directions, fading fast to nearly zero in most of the
non-equatorial-plane directions. This model is complicated enough
to cause that in the hemisphere facing the BH a quite complex
pattern emerges as a result of the  struggle between attraction
from gravitation and repulsion from plasma (see fig.
\ref{fig:disclike}). In both NSIS distributions, the effects of
plasma are negligible when $\omega_{obs}\gg\omega_{pl}(r_{obs})$
just like in the constant distribution.

The influence of plasma on the shadow of BH was previously studied
as well: analytic approach was cultivated in \cite{Perlick, Yan},
while numerical calculations were performed in \cite{Huang}.
Analytic approach was restricted to the specific class of plasma
distributions,
$\omega_{pl}^2=\rho^{-2}[f_r(r)+f_{\vartheta}(\vartheta)]$
\cite{Perlick}. Numerical calculations, not subject to this
constraint, produced the shape of the shadow with purely radial,
power law and exponential distributions. Out of these works, our
situation is most similar to \cite{Perlick}, where the authors
placed, like us, the observer at finite distance from the BH
($r=5$ in their case). All papers also included description of the
vacuum shadow, either directly or as a high frequency limit of the
plasma shadow. Near-extremal BH vacuum shadow exhibits a typical
feature -- flattened left hand side of the shadow, which was
present in all these papers and can be also seen in fig.
\ref{fig:vac}; this feature is called ``NHEKline'' in some papers,
e.g. in \cite{Yan}. In homogeneous plasma, the increase in the
size of the shadow with decreasing frequency was noticed in
\cite{Perlick}, and is also visible between figures \ref{fig:vac}
and \ref{fig:homog}. Paradoxically, distributions that decrease as
we are getting further from the BH and vanish as $r$ approaches
infinity have the opposite effect on the shadow: its size
decreases with decreasing frequency. This feature can be observed
in all three papers, as well as in figures \ref{fig:NSIS} and
\ref{fig:disclike}. A more detailed discussion of this topic can
be found in \cite{Perlick}. Finally, all figures, ours as well as
those in the papers \cite{Perlick, Huang, Yan}, suggest that
plasma rounds off the BH shadow and suppresses its typical
NHEKline signature at frequencies near $\omega_{pl}$.

Besides the plasma distribution, the form of SDF is influenced by
other circumstances such as BH rotation, observer's location and
observer's velocity. Rotation of black hole introduces asymmetry
between prograde and retrograde direction at any spatial point
outside the axis of rotation, and as a result, in case with no
plasma the rays evolving forward (backward) in time are pulled in
prograde (retrograde) direction. Thus, in our case the rays become
more squashed together in prograde direction and we obtain greater
SDF there (this is a naive opinion on the issue which seems to be
confirmed by numerical calculation). An observer boosted by a
significant fraction of speed of light in a certain direction sees
the light rays pulled towards each other in that direction due to
the special relativistic aberration of light, hence if the plasma
is absent, SDF rises there. Also the BH shadow is pulled in that
direction. These effects are modified in the presence of plasma.
The situation gets more complicated, since the evolution of rays
is now dependent on $\omega_{obs}$ in addition to their direction.
Therefore the set of rays used to calculate SDF for unboosted
observer does not coincide with the set of rays used to calculate
SDF for the boosted observer (this fact can be ignored in vacuum
due to achromaticity of the light bending). Thus, there is no
direct relation between SDF of two observers moving w.r.t. each
other. Let us just note that at large distances from the BH,
effects of gravitational pull are severely suppressed except for a
narrow region around the BH shadow, while the optical effects of
plasma may still be important depending on the plasma
distribution. However, in the distributions primarily concentrated
around the BH these effects are also suppressed everywhere except
nearby the BH shadow. On the other hand, if the observer is very
close to the BH, the gravitational pull dominates on a large
portion of the night sky, leaving only a small part of the sky
with well defined SDF that can be influenced by the presence of
plasma. The influence is not easy to predict, the analysis being
complicated by the fact that we do not have an analogue of Carter
constant for a generic plasma distribution, only for distributions
fulfilling the additional condition cited above, which was found
in \cite{Perlick} and further discussed or applied in calculations
in \cite{Cunha_review, BKT2017, Kimpson}.

\section{Conclusion}
We have discussed some theoretical aspects of general relativistic
geometrical optics in plasma, in particular, we have provided a
simple derivation of RDE which was, to our knowledge, not
presented in the literature before. We have also discussed
conservation laws and normalization constraints, writing them in a
covariant way. We have then introduced the concept of SDF and
calculated numerically the SDFs in various cases. To do that, we
have converted the EoMs for the reference ray, as well as the RDE,
into a set of first order ODE. This
allowed us to calculate the SDF point by point, in contrast to
constructing a network of reference rays and computing the
derivatives numerically, which would be certainly less exact,
especially for rays that undergo several revolutions around the BH
before escaping to infinity. The SDF was calculated for several
different plasma distributions: the vacuum (no plasma), the
homogeneous plasma, the NSIS and the flattened (disc-like) NSIS.
We have discussed also the dependence of SDF on some other
parameters, namely the distance of the observer from the BH, the
observational frequency and the spin of the BH. In future work
there are several directions which appear worth to follow. First,
it would be interesting to calculate the damping effect of plasma
on the light beam, the most dominant effect being probably the
Thompson scattering. Second, we would like to look at the SDF of
an observer outside the equatorial plane, which requires only
slight modification of the initial conditions. Finally, it would
be interesting to calculate the light curves of a distant star
seen by a distant observer on the other side of the BH, when the
incoming light is influenced simultaneously by the gravitational
field of the BH and the refraction index of the surrounding
plasma. Obviously, this is only interesting if the plasma itself
is not shining significantly, so that it does not outshine the
light from the distant star.

\section*{Acknowledgements}
The work was supported by the grant VEGA 1/0985/16. M. S. is also
thankful to the Comenius University for its support via the grants
UK/159/2018 and UK/105/2017.

\appendix
\section{Efficient variation procedure for the two-ray action}
\label{app:RDE}
\setcounter{equation}{0}
\renewcommand{\theequation}{A-\arabic{equation}}

Consider the action (\ref{Ray_action_1}) for a reference ray
$x^\mu(\lambda)$ and a neighboring ray $y^\mu(\lambda)$ lying in
the close vicinity of it. We can write
$y^\mu=x^\mu+\varepsilon\xi^\mu$, where $\varepsilon$ is a
positive parameter $\ll 1$ and $\bm\xi$ is the connecting vector.
Expand the action for the neighboring ray to the first order in
$\varepsilon$,
\begin{eqnarray}
S[y]\simeq S[x]+\varepsilon \int_{\lambda_1}^{\lambda_2}\left(\bm
k\cdot\nabla_{\bm k}\bm\xi - \frac{1}{2}\bm\xi\cdot \sharp
d\omega_{pl}^2\right) d\lambda\nonumber
\end{eqnarray}
When rewriting the integral, we started from the equality
\footnote{The equality just states that the connecting vector is
Lie-transported along the reference ray, $\mathcal{L}_{\bm k}\bm
\xi\vert_{x(\lambda)}=0$. The same formula is used in the
derivation of Jacobi equation.}$[\bm k,\bm \xi]=0$ at
$x(\lambda)$, which holds because $\bm \xi$ closes by definition a
4-sided polygon with $\bm k$, and we used its consequence, the
equality $\nabla_{\bm\xi}\bm k = \nabla_{\bm k}\bm\xi$, implied by
vanishing of the torsion tensor. Demanding both $x$ and $y$ to
obey the EoM for a ray suggests a variational principle for the
second term on the r.h.s., with the boundary values of both $x$
and $\bm\xi$ fixed. Thus, we can introduce the action for ray
deviation:
\begin{eqnarray}\label{Ray_action_2}
\psi [x,\bm\xi]=\int_{\lambda_1}^{\lambda_2}\left(\bm
k\cdot\nabla_{\bm k}\bm\xi - \frac{1}{2}\bm\xi\cdot \sharp
d\omega_{pl}^2\right)
d\lambda\equiv\int_{\lambda_1}^{\lambda_2}\mathcal{L} d\lambda
\end{eqnarray}
The first term is the action for the equation of geodesic
deviation a.k.a. Jacobi equation (similar term is proposed e.g. in \cite{Bazanski,
Bazanski2}), while the second term introduces plasma on the scene.
Let us write the $x$-variations as $x+\varepsilon\bm{\mathcal{X}}$
and the $\bm\xi$-variation as $\bm\xi+\varepsilon\bm\eta$, where
$\bm{\mathcal{X}},\bm\eta$ are vector fields vanishing at the
boundaries. Since we need to go only to the first order in
$\varepsilon$, we can discuss these variations separately. First
consider the $\bm\xi$-variation. It holds
\begin{eqnarray}
\mathcal{L}(x,\bm\xi+\varepsilon\bm\eta)-\mathcal{L}(x,\bm\xi)=
\varepsilon \left(\bm k\cdot\nabla_{\bm k}\bm\eta-\frac
12\bm\eta\cdot \sharp d\omega_{pl}^2\right)\hat =-\varepsilon
\bm\eta\cdot\left(\nabla_{\bm k}\bm k+\frac 12\sharp
d\omega_{pl}^2\right)\nonumber
\end{eqnarray}
where we introduced the effective equality $\hat =$ in which the
terms with the total derivative (terms of the form $\nabla_{\bm k}
(...)$) are dropped. The variation yields yet again the evolution
equation for the reference ray,
\begin{eqnarray}\label{app1:RE}
\nabla_{\bm k}\bm k=-\frac 12\sharp d\omega_{pl}^2
\end{eqnarray}
Now let us proceed to the variation w.r.t. $x$. We will use the
covariant variation rather than the standard one\footnote{The
difference is that the covariant variation will also cause a
change in $\bm\xi$, but this change is dependent on the
$x$-variation and will drop out eventually due to the ray
equation.}. Also, it is useful to realize that the commutation
relations allow us to swap $\nabla_{\bm k}\bm\xi$ for
$\nabla_{\bm\xi} \bm k$ and $\nabla_{\bm k}\bm{\mathcal{X}}$ for
$\nabla_{\bm{\mathcal{X}}}\bm k$. The geometric part of the
Lagrangian varies as follows:
\begin{eqnarray}
\begin{array}{ll}
\varepsilon^{-1}\delta_x\mathcal{L}_{geom}&=
\nabla_{\bm{\mathcal{X}}} (\bm k\cdot\nabla_{\bm k}\bm\xi)=
(\nabla_{\bm k}\bm{\mathcal{X}})
\cdot(\nabla_{\bm k}\bm\xi)+\bm k\cdot\nabla_{\bm{\mathcal{X}}}\nabla_{\bm k}\bm\xi\\
&\hat = -\bm{\mathcal{X}}\cdot\nabla_{\bm k}^2\bm\xi+\bm
k\cdot\nabla_{\bm k}
\nabla_{\bm{\mathcal{X}}}\bm\xi+\bm k\cdot R(\bm{\mathcal{X}},\bm k)\xi\\
&\hat = -\bm{\mathcal{X}}\cdot\nabla_{\bm k}^2\bm\xi-(\nabla_{\bm
k}\bm k)
\cdot(\nabla_{\bm{\mathcal{X}}}\bm\xi)+\bm{\mathcal{X}}\cdot R(\bm k,\bm\xi)\bm k\\
&\hat = -\bm{\mathcal{X}}\cdot[\nabla_{\bm k}^2\bm\xi-R(\bm
k,\bm\xi)\bm k+\bm A_{geom}]
\end{array}\nonumber
\end{eqnarray}
where $\bm A_{geom}$ is an auxiliary vector defined as $\bm
A_{geom}=\sharp[(\nabla_{\bm k}\bm k)\cdot(\nabla\bm\xi)]$. The
Riemann curvature operator $R(\bm U,\bm V)=\nabla_{\bm
U}\nabla_{\bm V}-\nabla_{\bm V}\nabla_{\bm U}-\nabla_{[\bm U,\bm
V]}$ was introduced in order to rearrange the covariant
derivatives, and in the course of calculation the identity $\bm
k\cdot R(\bm{\mathcal{X}},\bm k)\bm\xi=\bm{\mathcal{X}}\cdot R(\bm
k,\bm\xi)\bm k$, which follows from the symmetry of the Riemann
tensor w.r.t. the interchange of the pairs of indices, was used.
Let us now determine the $x$-variation of the plasma term:
\begin{eqnarray}
\varepsilon^{-1}\delta_x \mathcal{L}_{plasma}=-\frac{1}{2}
\nabla_{\bm{\mathcal{X}}}(\bm \xi\cdot\sharp d\omega_{pl}^2)
=-\bm{\mathcal{X}}\cdot\left[\frac{1}{2}\bm\xi\cdot
\nabla(\sharp
d\omega_{pl}^2)+\bm{A}_{plasma}\right]\nonumber
\end{eqnarray}
where $\bm{A}_{plasma}=\sharp[(1/2)(\sharp
d\omega_{pl}^2)\cdot(\nabla\bm\xi)]$. Collecting all terms we get:
\begin{eqnarray}
\varepsilon^{-1}\delta_x \mathcal{L}=-\bm{\mathcal{X}}\cdot
\left[\nabla_{\bm k}^2\bm\xi- R(\bm k,\bm\xi)\bm
k+\frac{1}{2}\sharp\nabla d\omega_{pl}^2(\bm\xi,.)+\bm
A\right]\nonumber
\end{eqnarray}
where we have isolated the auxiliary term $\bm
A=\bm{A}_{geom}+\bm{A}_{plasma}$, which appears as a consequence
of using the covariant variation and vanishes on the extremal due
to the ray equation (\ref{app1:RE}). Finally, using the definition
of the covariant Hessian matrix we arrive at the ray deviation
equation,
\begin{eqnarray}
\nabla_{\bm k}^2\bm\xi=R(\bm k,\bm\xi)\bm
k-\frac{1}{2}\sharp\bm{\mathcal{H}} (\bm\xi,.)
\end{eqnarray}

\section{Explicit formulae for RDE integration in Kerr geometry in
LNRF}\label{app:Kerr} \setcounter{equation}{0}
\renewcommand{\theequation}{B-\arabic{equation}}

\subsubsection*{Equations of motion for adjacent ray in LNRF}
Here we give the ``totally unpacked version'' of equations of
motion:
\begin{equation}\label{EoM_lnrf}
\begin{array}{ll}
\dot{r}&=(\sqrt{\Delta}/\rho)k^{\hat{r}}\\
\dot{\vartheta}&=(1/\rho)k^{\hat{\vartheta}}\\
\dot{k}^{\hat{r}}&=-\frac{1}{2}\mathcal{A}_{\hat{r}}-Kk^{\hat{r}}k^{\hat{\vartheta}}
-Lk^{\hat{\vartheta}}k^{\hat{\vartheta}}+Ik^{\hat{t}}k^{\hat{t}}-Mk^{\hat{\varphi}}
k^{\hat{\varphi}}+2Ok^{\hat{t}}k^{\hat{\varphi}}\\
\dot{k}^{\hat{\vartheta}}&=-\frac{1}{2}\mathcal{A}_{\hat{\vartheta}}+
Kk^{\hat{r}}k^{\hat{r}}+Lk^{\hat{r}}k^{\hat{\vartheta}}+Jk^{\hat{t}}k^{\hat{t}}-
Nk^{\hat{\varphi}}k^{\hat{\varphi}}+2Pk^{\hat{t}}k^{\hat{\varphi}}\\
\bullet k^{\hat{t}}&=(1/\alpha)(\varepsilon-\omega\mathcal{L})\\
\bullet k^{\hat{\varphi}}&=\mathcal{L}/\bar{\omega}\\
\dot{\xi}^{\hat{t}}&=\mathcal{V}^{\hat{t}}+O(k^{\hat{r}}\xi^{\hat{\varphi}}+k^{\hat{\varphi}}\xi^{\hat{r}})+
P(k^{\hat{\vartheta}}\xi^{\hat{\varphi}}+k^{\hat{\varphi}}\xi^{\hat{\vartheta}})+Ik^{\hat{t}}\xi^{\hat{r}}+
Jk^{\hat{t}}\xi^{\hat{\vartheta}}\\
\dot{\xi}^{\hat{\varphi}}&=\mathcal{V}^{\hat{\varphi}}+O(k^{\hat{r}}\xi^{\hat{t}}-k^{\hat{t}}\xi^{\hat{r}})+
P(k^{\hat{\vartheta}}\xi^{\hat{t}}-k^{\hat{t}}\xi^{\hat{\vartheta}})+Mk^{\hat{\varphi}}\xi^{\hat{r}}+
Nk^{\hat{\varphi}}\xi^{\hat{\vartheta}}\\
\dot{\xi}^{\hat{r}}&=\mathcal{V}^{\hat{r}}+O(k^{\hat{\varphi}}\xi^{\hat{t}}+k^{\hat{t}}\xi^{\hat{\varphi}})+
Ik^{\hat{t}}\xi^{\hat{t}}-Mk^{\hat{\varphi}}\xi^{\hat{\varphi}}-Kk^{\hat{r}}\xi^{\hat{\vartheta}}-
Lk^{\hat{\vartheta}}\xi^{\hat{\vartheta}}\\
\dot{\xi}^{\hat{\vartheta}}&=\mathcal{V}^{\hat{\vartheta}}+P(k^{\hat{\varphi}}\xi^{\hat{t}}+k^{\hat{t}}\xi^{\hat{\varphi}})+
Jk^{\hat{t}}\xi^{\hat{t}}-Nk^{\hat{\varphi}}\xi^{\hat{\varphi}}+Kk^{\hat{r}}\xi^{\hat{r}}+Lk^{\hat{\vartheta}}\xi^{\hat{r}}\\
\dot{\mathcal{V}}^{\hat{r}}&=-Q_1k^{\hat{\vartheta}}(k^{\hat{r}}\xi^{\hat{\vartheta}}-k^{\hat{\vartheta}}\xi^{\hat{r}})-
Q_2k^{\hat{\vartheta}}(k^{\hat{t}}\xi^{\hat{\varphi}}-k^{\hat{\varphi}}\xi^{\hat{t}})-Q_1(c_1k^{\hat{t}}-Sk^{\hat{\varphi}})
(k^{\hat{r}}\xi^{\hat{t}}-k^{\hat{t}}\xi^{\hat{r}})\\
&+Q_1(Sk^{\hat{t}}-c_2k^{\hat{\varphi}})(k^{\hat{r}}\xi^{\hat{\varphi}}-k^{\hat{\varphi}}\xi^{\hat{r}})+
Q_2(Sk^{\hat{t}}-c_2k^{\hat{\varphi}})(k^{\hat{\vartheta}}\xi^{\hat{t}}-k^{\hat{t}}\xi^{\hat{\vartheta}})\\
&-Q_2(c_1k^{\hat{t}}-Sk^{\hat{\varphi}})(k^{\hat{\vartheta}}\xi^{\hat{\varphi}}-k^{\hat{\varphi}}\xi^{\hat{\vartheta}})-
\frac{1}{2}\left(\mathcal{H}_{\hat{r}\hat{r}}\xi^{\hat{r}}+\mathcal{H}_{\hat{r}\hat{\vartheta}}
\xi^{\hat{\vartheta}}\right)\\
&+O(k^{\hat{\varphi}}\mathcal{V}^{\hat{t}}+k^{\hat{t}}\mathcal{V}^{\hat{\varphi}})+Ik^{\hat{t}}\mathcal{V}^{\hat{t}}-
Mk^{\hat{\varphi}}\mathcal{V}^{\hat{\varphi}}-Kk^{\hat{r}}\mathcal{V}^{\hat{\vartheta}}-Lk^{\hat{\vartheta}}
\mathcal{V}^{\hat{\vartheta}}\\
\dot{\mathcal{V}}^{\hat{\vartheta}}&=Q_1k^{\hat{r}}(k^{\hat{r}}\xi^{\hat{\vartheta}}-k^{\hat{\vartheta}}\xi^{\hat{r}})+
Q_2k^{\hat{r}}(k^{\hat{t}}\xi^{\hat{\varphi}}-k^{\hat{\varphi}}\xi^{\hat{t}})+Q_2(Sk^{\hat{t}}-c_1k^{\hat{\varphi}})
(k^{\hat{r}}\xi^{\hat{t}}-k^{\hat{t}}\xi^{\hat{r}})\\
&-Q_2(c_2k^{\hat{t}}-Sk^{\hat{\varphi}})(k^{\hat{r}}\xi^{\hat{\varphi}}-k^{\hat{\varphi}}\xi^{\hat{r}})+Q_1(c_2k^{\hat{t}}-
Sk^{\hat{\varphi}})(k^{\hat{\vartheta}}\xi^{\hat{t}}-k^{\hat{t}}\xi^{\hat{\vartheta}})\\
&-Q_1(Sk^{\hat{t}}-c_1k^{\hat{\varphi}})(k^{\hat{\vartheta}}\xi^{\hat{\varphi}}-k^{\hat{\varphi}}\xi^{\hat{\vartheta}})-
\frac{1}{2}\left(\mathcal{H}_{\hat{\vartheta}\hat{r}}\xi^{\hat{r}}+\mathcal{H}_{\hat{\vartheta}\hat{\vartheta}}
\xi^{\hat{\vartheta}}\right)\\
&+P(k^{\hat{\varphi}}\mathcal{V}^{\hat{t}}+k^{\hat{t}}\mathcal{V}^{\hat{\varphi}})+Jk^{\hat{t}}\mathcal{V}^{\hat{t}}-
Nk^{\hat{\varphi}}\mathcal{V}^{\hat{\varphi}}+Kk^{\hat{r}}\mathcal{V}^{\hat{r}}+Lk^{\hat{\vartheta}}\mathcal{V}^{\hat{r}}\\
\bullet\mathcal{V}^{\hat{\varphi}}&=
(1/\bar{\omega})P_\varphi-M(k^{\hat{r}}\xi^{\hat{\varphi}}-k^{\hat{\varphi}}\xi^{\hat{r}})-
N(k^{\hat{\vartheta}}\xi^{\hat{\varphi}}-k^{\hat{\varphi}}\xi^{\hat{\vartheta}})-O(k^{\hat{t}}\xi^{\hat{r}}-k^{\hat{r}}\xi^{\hat{t}})-
P(k^{\hat{t}}\xi^{\hat{\vartheta}}-k^{\hat{\vartheta}}\xi^{\hat{t}})\\
\bullet\mathcal{V}^{\hat{t}}&=
(1/k^{\hat{t}})\left[\mathcal{V}^{\hat{r}}k^{\hat{r}}+\mathcal{V}^{\hat{\vartheta}}
k^{\hat{\vartheta}}+\mathcal{V}^{\hat{\varphi}}k^{\hat{\varphi}}+\frac{1}{2}(\xi^{\hat{r}}\mathcal{A}^{\hat{r}}+
\xi^{\hat{\vartheta}}\mathcal{A}^{\hat{\vartheta}})\right]\nonumber
\end{array}
\end{equation}
The equations marked with the bullet are algebraic rather than
differential and come from the conservation laws. The unknown
symbols on the right hand sides encode Kerr geometry and plasma
distribution and are defined in the following two paragraphs.

The complete set of equations comprises 10 differential and 4
algebraic equations which are to be solved twice for every
reference ray (since the ellipse is generated by two vectors
${\bm{\xi}}_{(k)}$)\footnote{Alternatively, we can solve $6 +16$
equations simultaneously, should it prove to be faster. In fact,
this is how we proceeded.}. There are also optional evolution
equations for temporal and azimuthal coordinate,
$\dot{t}=g^{tA}k_A$ and $\dot{\varphi}=g^{\varphi A}k_A$, which
may be added to the list to obtain the complete set of 16
equations of motion. Note, that Boyer-Lindquist coordinates are
ill-defined on north/south pole. This can cause problems, if we
try to integrate the system of rays with the reference ray
crossing one of the poles (rays with $\mathcal{L}=0$). The
equations for reference ray cope with this problem by letting
$\vartheta$ escape from the interval $[0,\pi]$, but the equations
for neighbouring ray blow up because of term
$P_\varphi/\bar{\omega}$, so it is generally good idea to avoid
integrating these rays when calculating SDF. Rays closely passing
the poles may also occasionally suffer from numerical problems,
see e.g. blue spots on fig. \ref{fig:disclike} near values
$\bar\varphi=\pi/2, 3\pi/2$.

\subsubsection*{Kerr geometry}
Kerr metric in the Boyer--Lindquist coordinates, in the notations
adopted from Thorne et al. \cite{Thorne}, is:
\begin{eqnarray}\label{Metric_tensor}
ds^2=-\alpha^2 dt^2 + \bar{\omega}^2(d\varphi-\omega
dt)^2+\frac{\rho^2}{\Delta}dr^2+\rho^2d\vartheta^2
\end{eqnarray}
where $\Delta = r^2 - 2r + a^2$, $\rho^2 = r^2 + a^2 \cos^2
\theta$ and
\begin{eqnarray}
\alpha=\frac{\rho\sqrt{\Delta}}{\Sigma}\qquad
\omega=\frac{2ar}{\Sigma^2}\qquad\bar\omega=\frac{\Sigma\sin\vartheta}{\rho}
\qquad \Sigma^2=(r^2+a^2)^2-a^2\Delta\sin^2\vartheta\nonumber
\end{eqnarray}
In this form an orthonormal covector basis
$(\bm{\omega}^{\hat{t}}, \bm{\omega}^{\hat{t}},
\bm{\omega}^{\hat{\vartheta}}, \bm{\omega}^{\hat{\varphi}})$
arises naturally. The corresponding vector basis
$(\bm{e}_{\hat{t}}, \bm{e}_{\hat{r}}, \bm{e}_{\hat{\vartheta}},
\bm{e}_{\hat{\varphi}})$, carried by zero angular momentum
observers, is called locally non-rotating reference frame (LNRF).
The two bases are:
\begin{eqnarray}\label{lnrf_to_coordinate}
\left\lbrace\begin{array}{ll}
\bm{\omega}^{\hat{t}}&=\alpha \bm dt\\
\bm{\omega}^{\hat{r}}&=(\rho/\sqrt{\Delta}) \bm dr\\
\bm{\omega}^{\hat{\vartheta}}&=\rho \bm d\vartheta\\
\bm{\omega}^{\hat{\varphi}}&=\bar{\omega}(\bm d\varphi-\omega \bm dt)\\
\end{array}\right.
\qquad\qquad\qquad \left\lbrace\begin{array}{ll}
\bm{e}_{\hat{t}}&=(1/\alpha) (\partial_t+\omega\partial_\varphi)\\
\bm{e}_{\hat{r}}&=(\sqrt{\Delta}/\rho) \partial_r\\
\bm{e}_{\hat{\vartheta}}&=(1/\rho) \partial_\vartheta\\
\bm{e}_{\hat{\varphi}}&=(1/\bar{\omega}) \partial_\varphi\\
\end{array}\right.
\end{eqnarray}
Connection forms $\bm{\omega}_{\hat{\mu}\hat{\nu}}$
($(\omega_{\hat{\mu}\hat{\nu}})_{\hat{\lambda}}=\Gamma_{\hat{\mu}\hat{\nu}\hat{\lambda}}$,
where hatted indices refer to LNRF) are:
\begin{eqnarray}
\begin{array}{lll}
\bm \omega_{\hat{t}\hat{r}}=&I\bm e^{\hat{t}}&+O\bm e^{\hat{\varphi}}\\
\bm \omega_{\hat{t}\hat{\vartheta}}=&J\bm e^{\hat{t}}&+P\bm e^{\hat{\varphi}}\\
\end{array}
\qquad\qquad
\begin{array}{lll}
\bm \omega_{\hat{t}\hat{\varphi}}=&O\bm e^{\hat{r}}&+P\bm e^{\hat{\vartheta}}\\
\bm \omega_{\hat{r}\hat{\vartheta}}=&K\bm e^{\hat{r}}&+L\bm e^{\hat{\vartheta}}\\
\end{array}
\qquad\qquad
\begin{array}{lll}
\bm \omega_{\hat{r}\hat{\varphi}}=&-O\bm e^{\hat{t}}&+M\bm e^{\hat{\varphi}}\\
\bm \omega_{\hat{\vartheta}\hat{\varphi}}=&-P\bm e^{\hat{t}}&+N\bm e^{\hat{\varphi}}\\
\end{array}\nonumber
\end{eqnarray}
where $\omega_{\hat{\nu}\hat{\mu}}= -\omega_{\hat{\mu}\hat{\nu}}$
and
\begin{eqnarray}
\begin{array}{lll}
I=-\dfrac{\alpha_{,r}}{\alpha}\dfrac{\sqrt{\Delta}}{\rho}&=-
\dfrac{1}{\rho^3\sqrt{\Delta}\Sigma^2}\{r\Delta\Sigma^2+\Sigma^2\rho^2(r-1)-\rho^2
\Delta[2r(r^2+a^2)-a^2s^2(r-1)]\}\\
J=-\dfrac{\alpha_{,\vartheta}}{\alpha}\dfrac{1}{\rho}&=
\dfrac{sca^2}{\rho^3\Sigma^2}(\Sigma^2-\rho^2 \Delta)\\
K=\dfrac{\rho_{,\vartheta}}{\rho^2}&=-\dfrac{a^2sc}{\rho^3}\\
L=-\dfrac{\rho_{,r}\sqrt{\Delta}}{\rho^2}&=-\dfrac{r\sqrt{\Delta}}{\rho^3}\\
M=-\dfrac{\bar{\omega}_{,r}}{\bar{\omega}}
\dfrac{\sqrt{\Delta}}{\rho}&=
-\frac{\sqrt{\Delta}}{\Sigma^2\rho^3}\{\rho^2[2r(r^2+a^2)-a^2s^2(r-1)]-r\Sigma^2\}\\
N=-\dfrac{\bar{\omega}_{,\vartheta}}{\bar{\omega}}
\dfrac{1}{\rho}&=
-\dfrac{c}{s\Sigma^2\rho^3}(\Sigma^2\rho^2-a^2s^2\rho^2\Delta +a^2s^2\Sigma^2)\\
O=-\dfrac{\bar{\omega}\omega_{,r}\sqrt{\Delta}}{2\alpha\rho}&=
-\dfrac{as}{\Sigma^2\rho^3}[\Sigma^2 -4r^2(r^2+a^2)+2a^2s^2r(r-1)]\\
P=-\dfrac{\bar{\omega}\omega_{,\vartheta}}{2\alpha\rho}&= -
\dfrac{2a^3s^2cr\sqrt{\Delta}}{\Sigma^2\rho^3}
\end{array}\nonumber
\end{eqnarray}
We have used the abbreviations $s=\sin\vartheta$ and
$c=\cos\vartheta$. The formulae can be looked up in \cite{BPT} or
obtained directly by solving Cartan's structure equations (to be
found e.g. in \cite{Fecko}).

The LNRF curvature tensor
$R_{\hat{\mu}\hat{\nu}\hat{\kappa}\hat{\lambda}}$, as computed in
\cite{BPT}, is
\begin{eqnarray}
R_{\hat{\mu}\hat{\nu}\hat{\kappa}\hat{\lambda}} = \left(
\begin{array}{cccccc}
 -c_1Q_1 & SQ_2 & 0 & 0 & -SQ_1 & c_1Q_2 \\
 & c_2Q_1 & 0 & 0 & c_2Q_2 & SQ_1 \\
 & & Q_1 & -Q_2 & 0 & 0 \\
 & & & -Q_1 & 0 & 0 \\
 & & & & -c_2Q_1 & SQ_2\\
 & & & & & c_1Q_1
\end{array} \right)\nonumber
\end{eqnarray}
where the pairs of indices $(\hat{\mu}\hat{\nu})$ and
$(\hat{\kappa}\hat{\lambda})$ assume values $(\hat{t}\hat{r})$,
$(\hat{t}\hat{\vartheta})$, $(\hat{t}\hat{\varphi})$,
$(\hat{r}\hat{\vartheta})$, $(\hat{r}\hat{\varphi})$ and
$(\hat{\vartheta}\hat{\varphi})$ (in this order), and
\begin{eqnarray}
\begin{array}{lll}
Q_1=\dfrac{r(r^2-3a^2c^2)}{\rho^6} \qquad\qquad
&Q_2=\dfrac{ac(3r^2-a^2c^2)}{\rho^6} \qquad\qquad
&S=\dfrac{3as\sqrt{\Delta}(r^2+a^2)}{\Sigma^2}\\
c_1=\dfrac{2+z}{1-z} \qquad\qquad &c_2=\dfrac{1+2z}{1-z}
\qquad\qquad &z=\dfrac{\Delta a^2s^2}{(r^2+a^2)^2}
\end{array}\nonumber
\end{eqnarray}
The matrix is symmetric because of the symmetry of
$R_{\hat{\mu}\hat{\nu}\hat{\kappa}\hat{\lambda}}$ in the pairs of
indices.

\subsubsection*{Plasma description}
In our calculations we have restricted ourselves to stationary,
axially symmetric plasma distributions,
$\omega_{pl}^2=\omega_{pl}^2(r,\vartheta)$. Gradient of the
function $\omega_{pl}^2$ is
$d\omega_{pl}^2=\mathcal{A}_{\hat{r}}\bm{e^{\hat{r}}}+\mathcal{A}_{\hat{\vartheta}}\bm{e^{\hat{\vartheta}}}$,
where $\mathcal{A}_{\hat{r}}=
(\sqrt{\Delta}/\rho)(\omega_{pl}^2)_{,r}$ and
$\mathcal{A}_{\hat{\vartheta}}=(1/\rho)(\omega_{pl}^2)_{,\vartheta}$.
The Hessian matrix can be easily calculated using the formula
\begin{eqnarray}
\mathcal{H}_{\hat{\mu}\hat{\nu}}=(\nabla
d\omega_{pl}^2)_{\hat{\mu}\hat{\nu}}=\langle\nabla_{\hat{\nu}}
d\omega_{pl}^2,\bm{e_{\hat{\mu}}}\rangle=\nabla_{\hat{\nu}}\langle
d\omega_{pl}^2,\bm{e_{\hat{\mu}}}\rangle -\langle
d\omega_{pl}^2,\nabla_{\hat{\nu}}\bm{e_{\hat{\mu}}}\rangle\nonumber
\end{eqnarray}
The matrix is symmetric and has only 6 nonzero components:
\begin{eqnarray}
\begin{array}{ll}
\mathcal{H}_{\hat{t}\hat{t}}=&I\mathcal{A}_{\hat{r}}+J\mathcal{A}_{\hat{\vartheta}}\\
\mathcal{H}_{\hat{r}\hat{r}}=&(\sqrt{\Delta}/\rho)\mathcal{A}_{\hat{r},r}+ K\mathcal{A}_{\hat{\vartheta}}\\
\mathcal{H}_{\hat{\vartheta}\hat{\vartheta}}=&(1/\rho)\mathcal{A}_{\hat{\vartheta},\vartheta}-L\mathcal{A}_{\hat{r}}\\
\end{array}
\qquad\qquad\qquad
\begin{array}{ll}
\mathcal{H}_{\hat{\varphi}\hat{\varphi}}=&-M\mathcal{A}_{\hat{r}}-N\mathcal{A}_{\hat{\vartheta}}\\
\mathcal{H}_{\hat{t}\hat{\varphi}}=&O\mathcal{A}_{\hat{r}}+P\mathcal{A}_{\hat{\vartheta}}\\
\mathcal{H}_{\hat{\vartheta}\hat{r}}=&(\sqrt{\Delta}/\rho)\mathcal{A}_{\hat{\vartheta},r}- K\mathcal{A}_{\hat{r}}\\
\end{array}\nonumber
\end{eqnarray}

\section{Numerical results}\label{app:results}
To illustrate the general features of the star distribution
function discussed in the text, let us present here some example
figures with our results.

\begin{figure}[H]
\includegraphics[scale=0.3]{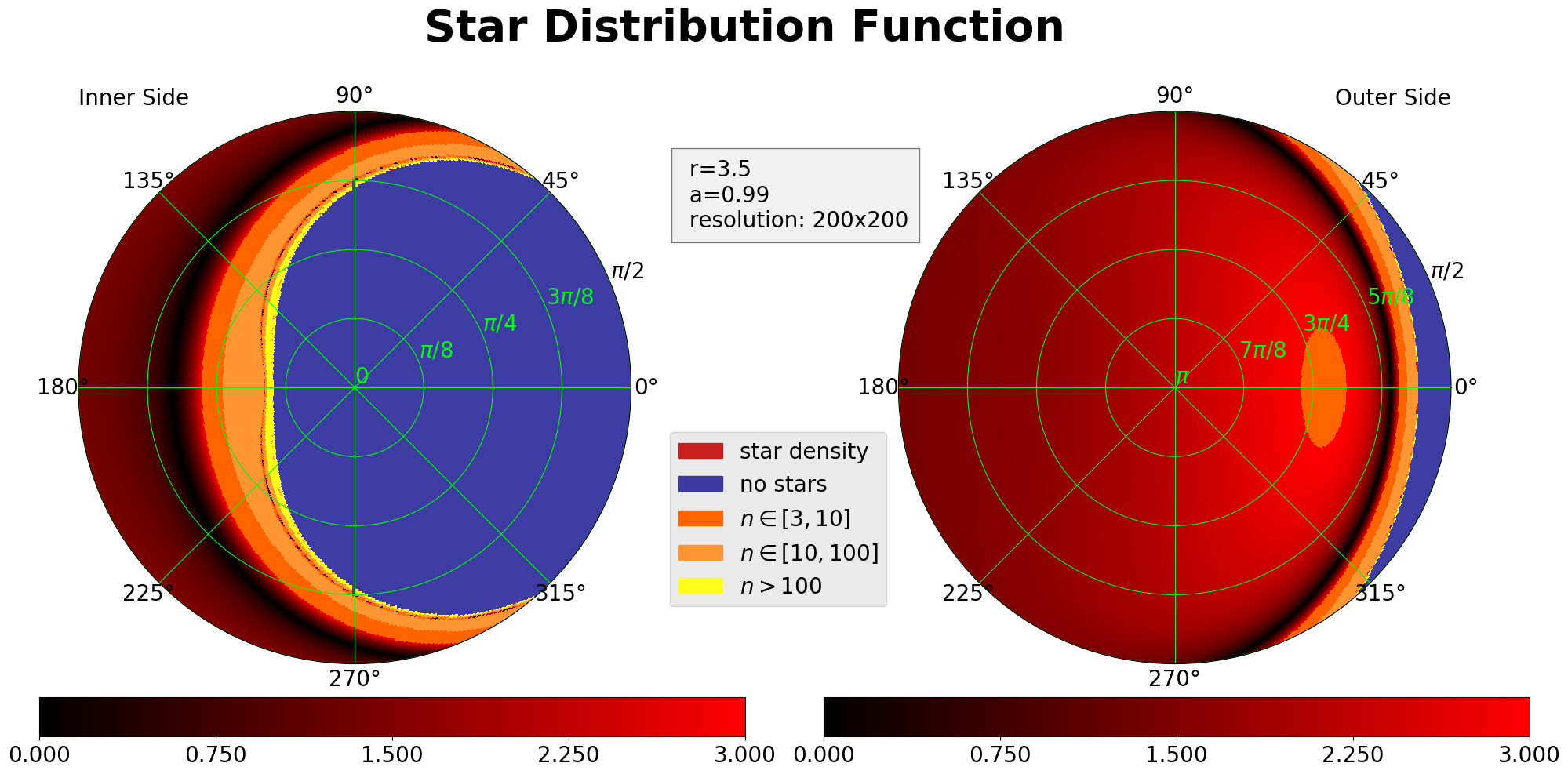}
\centering \caption{SDF (star distribution function introduced in
(\ref{SDF_def}), which has been calculated from
(\ref{SDF_num_formula}) and is denoted in the figure by $n$), as
measured by an observer at rest with respect to LNRF (locally
non-rotating reference frame). The SDF was computed for vacuum,
observer's radial coordinate $r$=3.5 and Kerr parameter $a$=0.99.
The resolution indicates the size of the grid in $\bar\vartheta$
and $\bar\varphi$ direction, on which the SDF was calculated. The
two circles represent two halves of the local sky, the half
looking roughly towards the BH (inner side) and the half looking
roughly away from the BH (outer side). Inside the circles, the
coordinate $\bar\varphi$ varies in azimuthal direction and the
coordinate $\bar\vartheta$ varies in radial direction (and scales
linearly)} \label{fig:vac}
\end{figure}

\begin{figure}[H]
\includegraphics[scale=0.3]{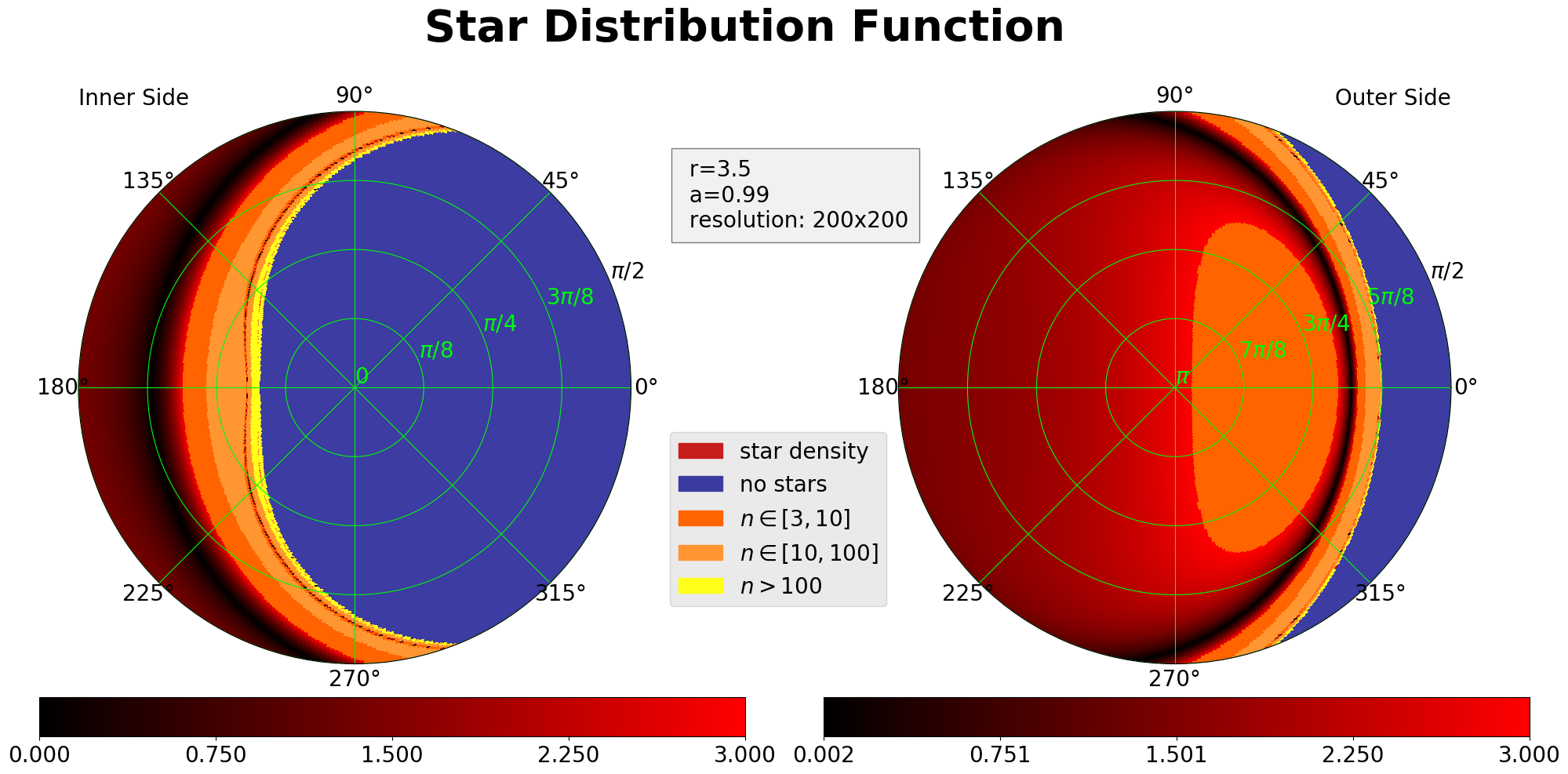}
\centering \caption{SDF as measured by the LNRF observer in
plasma, $r$=3.5, $a$=0.99, ratio of observational and plasma
frequency $\omega_{obs}/\omega_{pl}=2$, plasma distribution is
homogeneous (the value of $\omega_{pl}$ can be scaled away)}
\label{fig:homog}
\end{figure}

\begin{figure}[H]
\includegraphics[scale=0.3]{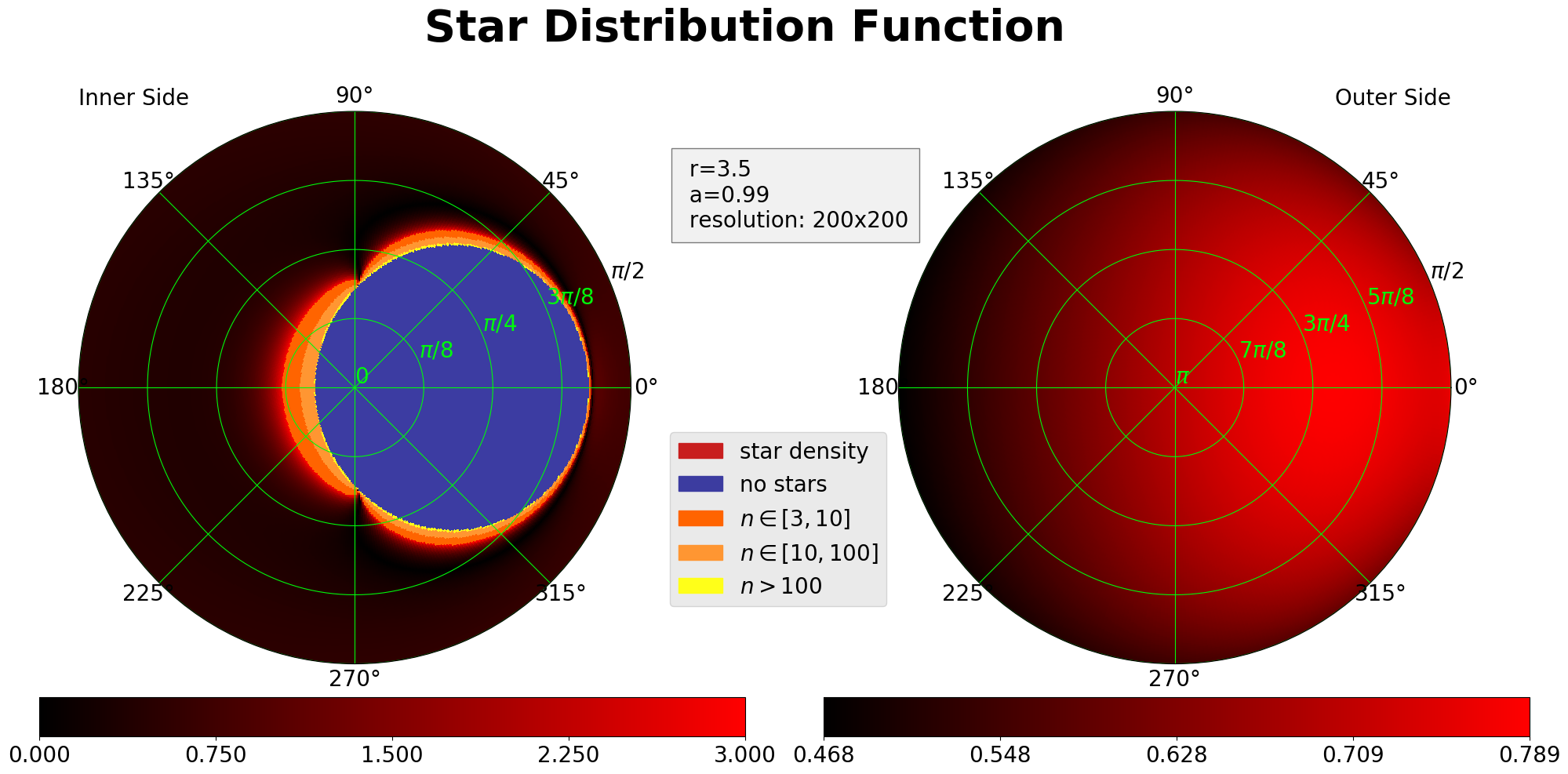}
\centering \caption{SDF as measured by the LNRF observer in
plasma, $r$=3.5, $a$=0.99, $\omega_{obs}/\omega_{pl}=1.1$, plasma
distribution is NSIS (non-singular isothermal sphere) with $r_c=1$
(the value of $K$ can be scaled away)} \label{fig:NSIS}
\end{figure}

\begin{figure}[H]
\includegraphics[scale=0.3]{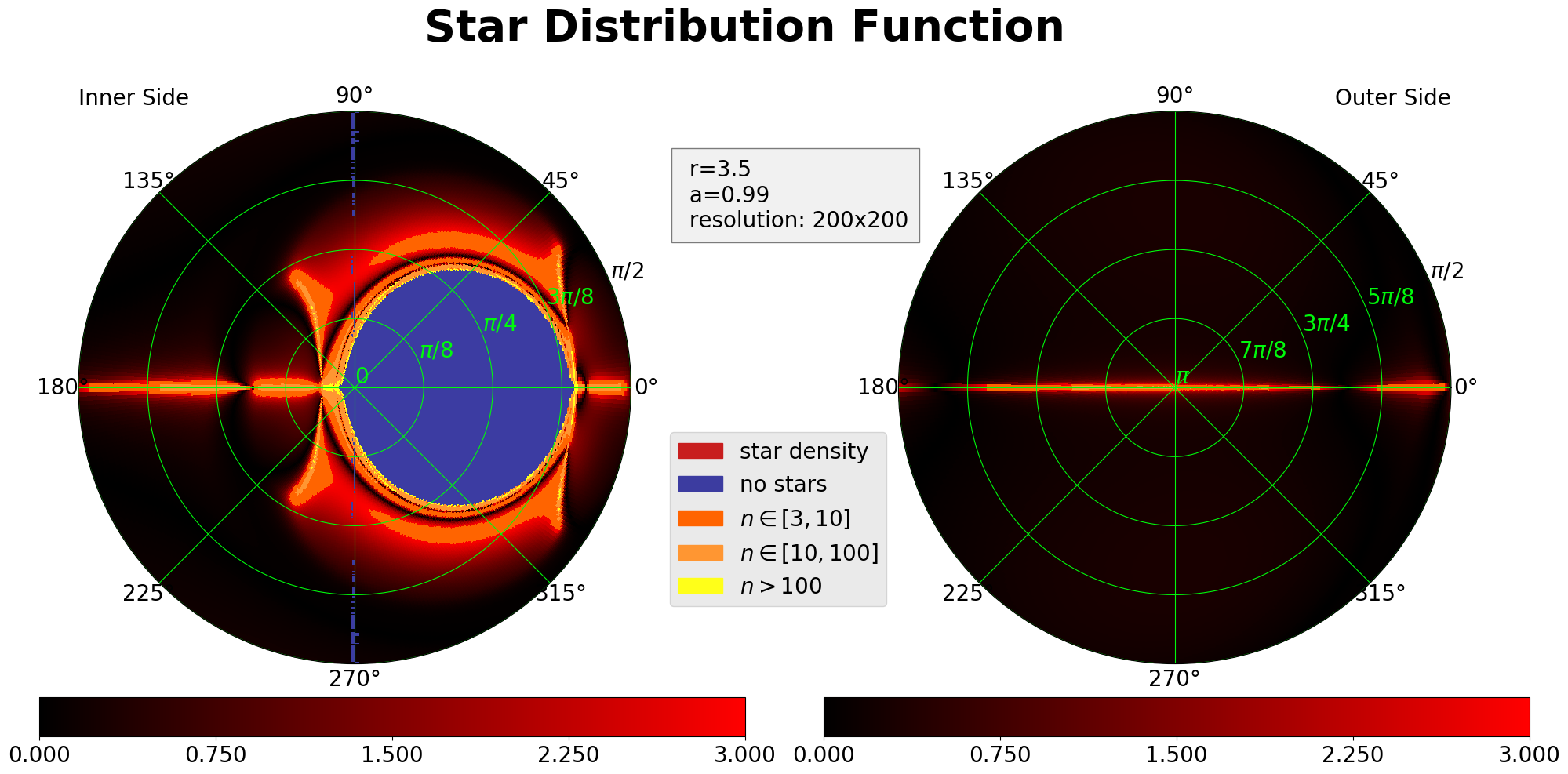}
\centering \caption{SDF as measured by the LNRF observer in plasma,
$r$=3.5, $a$=0.99, $\omega_{obs}/\omega_{pl}=1.1$, plasma
distribution is flattened NSIS with $s=0.1$, $r_c=1$ (the value of $K$ can be scaled away)}
\label{fig:disclike}
\end{figure}

\begin{figure}[H]
\includegraphics[scale=0.8]{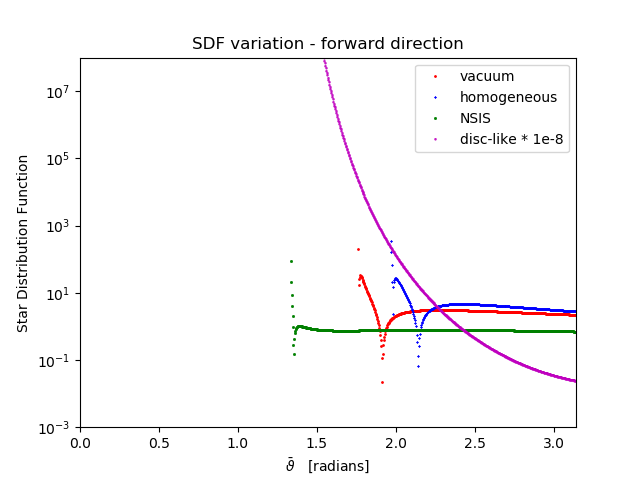}
\centering \caption{SDF with fixed direction $\bar\varphi=0$
(``forward'' w.r.t. the BH rotation), for cases depicted in
figures \ref{fig:vac}, \ref{fig:homog}, \ref{fig:NSIS},
\ref{fig:disclike}} \label{fig:var_fwd}
\end{figure}

\begin{figure}[H]
\includegraphics[scale=0.8]{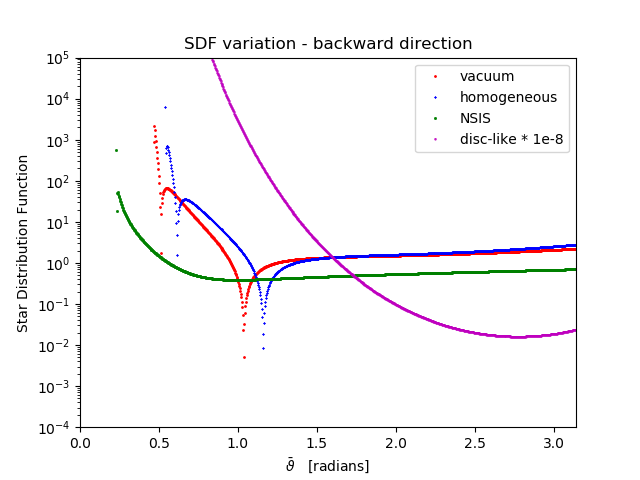}
\centering \caption{SDF with fixed direction $\bar\varphi=\pi$,
for cases depicted in figures \ref{fig:vac}, \ref{fig:homog},
\ref{fig:NSIS}, \ref{fig:disclike}. The functional dependence is
terminated upon reaching the BH shadow.} \label{fig:var_bwd}
\end{figure}

\begin{figure}[H]
\includegraphics[scale=0.8]{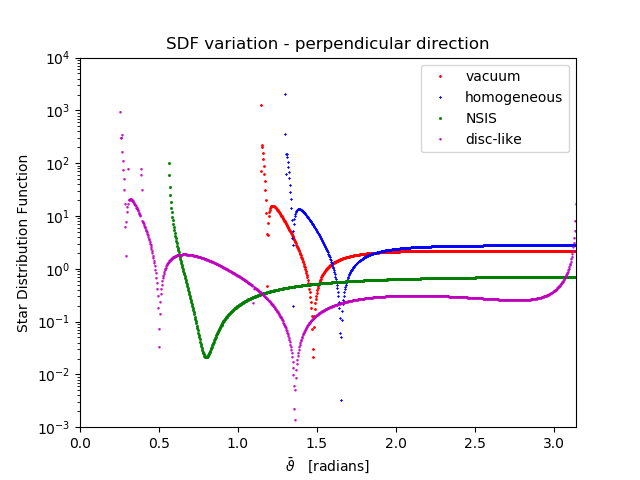}
\centering \caption{SDF with fixed direction $\bar\varphi=\pi/2$,
for cases depicted in figures \ref{fig:vac}, \ref{fig:homog},
\ref{fig:NSIS}, \ref{fig:disclike}}
\label{fig:var_down}
\end{figure}

\begin{figure}[H]
\begin{center}
\begin{multicols}{2}
    \includegraphics[width=\linewidth]{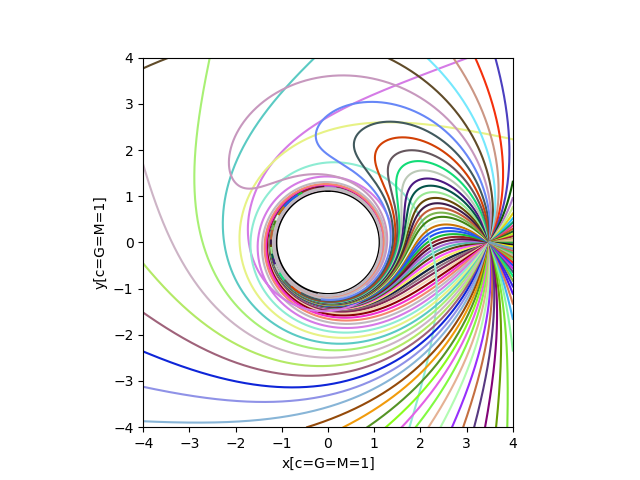}
    a)\\
    \includegraphics[width=\linewidth]{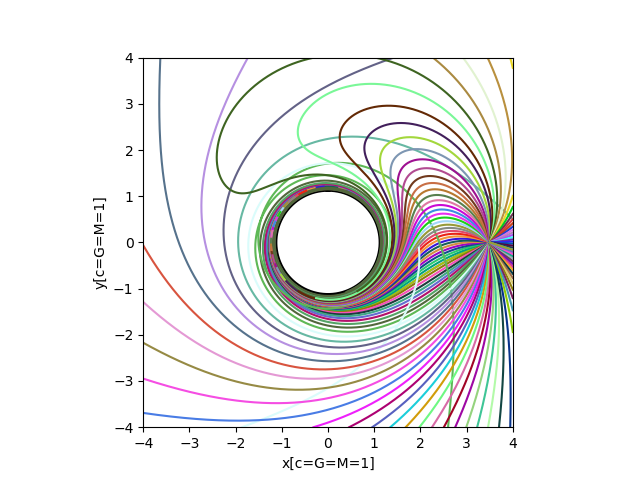}
    b)\par
    \includegraphics[width=\linewidth]{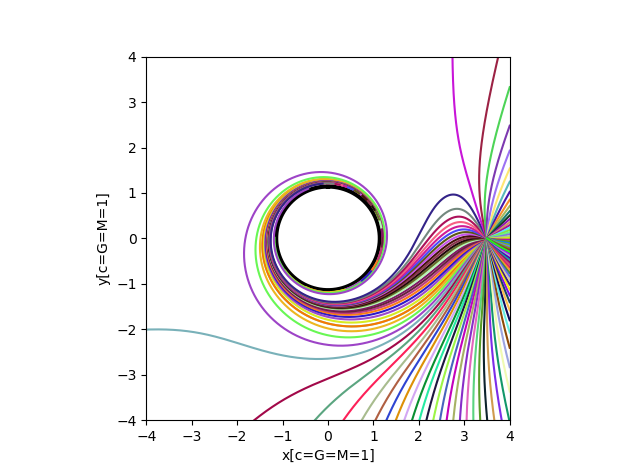}
    c)\\
    \includegraphics[width=\linewidth]{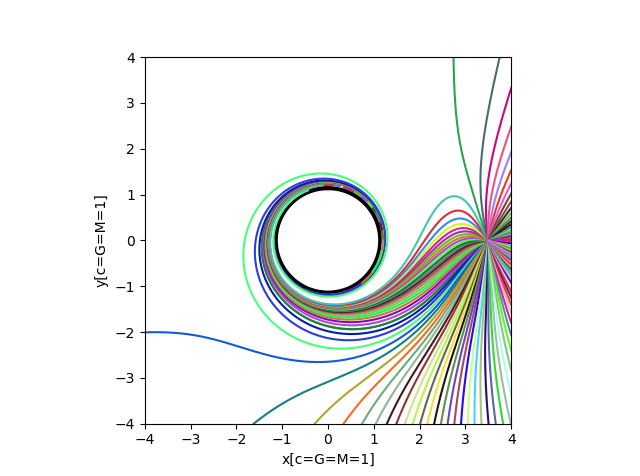}
    d)\\
\end{multicols}
\caption{Backward tracing of rays in equatorial plane for cases
depicted in figures \ref{fig:vac}, \ref{fig:homog},
\ref{fig:NSIS}, \ref{fig:disclike}. The projection treats
coordinates as if they were Euclidean. The plasma distribution
models are: a) vacuum, b) homogeneous, c) NSIS d) disc-like.}
\label{fig:equatorial_raytrace}
\end{center}
\end{figure}

\begin{figure}[H]
\begin{center}
\begin{multicols}{2}
    \includegraphics[width=\linewidth]{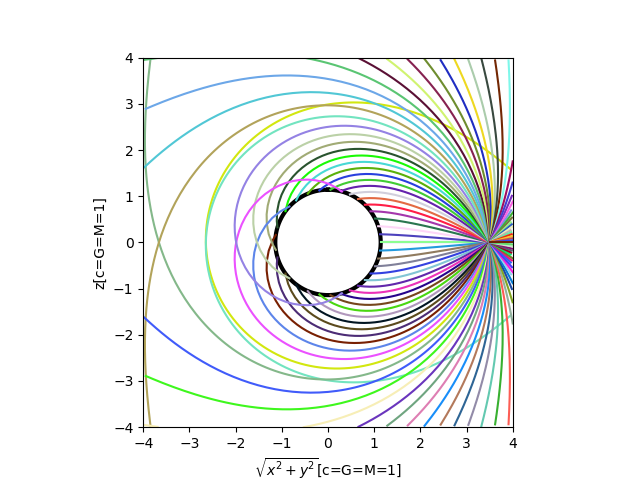}
    a)\\
    \includegraphics[width=\linewidth]{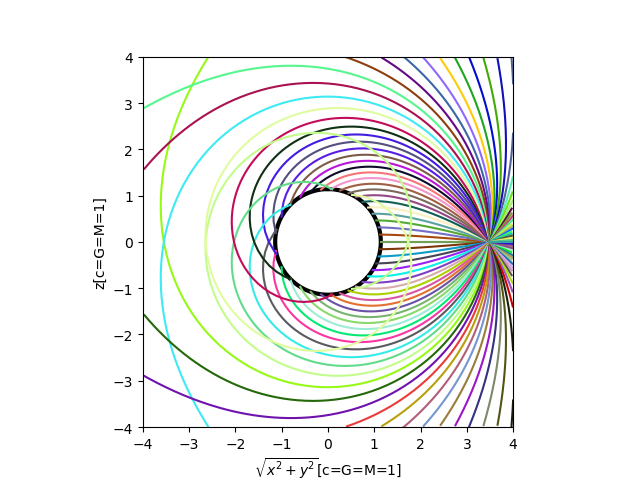}
    b)\par
    \includegraphics[width=\linewidth]{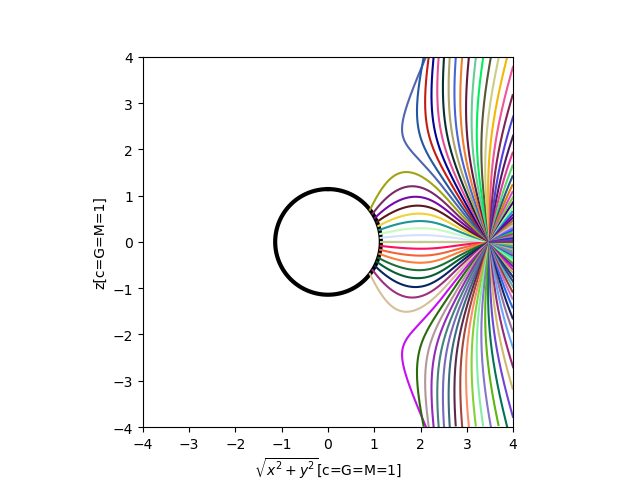}
    c)\\
    \includegraphics[width=\linewidth]{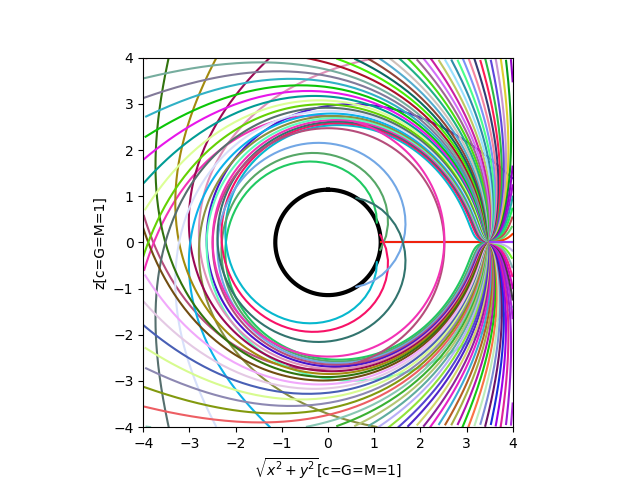}
    d)\\
\end{multicols}
\caption{Backward tracing of rays in the plane drawn through the
axis of BH for cases depicted in figures \ref{fig:vac},
\ref{fig:homog}, \ref{fig:NSIS}, \ref{fig:disclike}. The
projection treats coordinates as if they were Euclidean. The
azimuthal dependence is not depicted. The plasma distribution
models are: a) vacuum, b) homogeneous, c) NSIS d) disc-like.
Negative values on the horizontal axis indicate that the ray has
passed through south/north pole. } \label{fig:vertical_raytrace}
\end{center}
\end{figure}

\end{document}